\documentclass[letterpaper]{article} 
\usepackage{aaai25}  
\usepackage{times}  
\usepackage{helvet}  
\usepackage{courier}  
\usepackage[hyphens]{url}  
\usepackage{graphicx} 
\usepackage{natbib}  
\usepackage{caption} 
\usepackage{algorithm}
\usepackage{algorithmic}

\usepackage{graphicx} 

\usepackage{amsmath}
\newtheorem{theorem}{Theorem}
\newtheorem{assumption}{Assumption}

\usepackage{amssymb}
\usepackage{multirow}
\usepackage{makecell}
\usepackage{colortbl} 

\usepackage{arydshln}

\definecolor{colorFst}{rgb}{0.752,0.89,0.792}          
\definecolor{colorSnd}{rgb}{0.886,0.929,0.725}      
\definecolor{colorTrd}{rgb}{1,0.980,0.757}     

\newcommand{\fs}{\bf}   

\usepackage{newfloat}
\usepackage{listings}
\DeclareCaptionStyle{ruled}{labelfont=normalfont,labelsep=colon,strut=off} 
\floatstyle{ruled}
\newfloat{listing}{tb}{lst}{}
\floatname{listing}{Listing}
\title{Representing Sounds as Neural Amplitude Fields: A Benchmark of Coordinate-MLPs and A Fourier Kolmogorov-Arnold Framework}

\author {
   Linfei Li\textsuperscript{\rm 1},
    Lin Zhang\textsuperscript{\rm 1}\thanks{Corresponding Author},
    Zhong Wang\textsuperscript{\rm 2},
    Fengyi Zhang\textsuperscript{\rm 3},
    Zelin Li\textsuperscript{\rm 4},
    Ying Shen\textsuperscript{\rm 1}
}
\affiliations {
    \textsuperscript{\rm 1}School of Computer Science and Technology, Tongji University\\
    \textsuperscript{\rm 2} Department of Automation, Shanghai Jiao Tong University\\
    \textsuperscript{\rm 3} School of Electrical Engineering and Computer Science,  The University of Queensland \\
    \textsuperscript{\rm 4}McCormick School of Engineering, Northwestern University \\
    cslinfeili@tongji.edu.cn, cslinzhang@tongji.edu.cn, cszhongwang@sjtu.edu.cn, \\ fengyi.zhang@uq.edu.au, zelinli2025@u.northwestern.edu, yingshen@tongji.edu.cn
}
\usepackage{bibentry}

\begin{document}

\maketitle

\begin{abstract}
Although Coordinate-MLP-based implicit neural representations have excelled in representing radiance fields, 3D shapes, and images, their application to audio signals remains underexplored. To fill this gap, we investigate existing implicit neural representations, from which we extract 3 types of positional encoding and 16 commonly used activation functions. Through combinatorial design, we establish the first benchmark for Coordinate-MLPs in audio signal representations. Our benchmark reveals that Coordinate-MLPs require complex hyperparameter tuning and frequency-dependent initialization, limiting their robustness. To address these issues, we propose Fourier-ASR, a novel framework based on the Fourier series theorem and the Kolmogorov-Arnold representation theorem. Fourier-ASR introduces Fourier Kolmogorov-Arnold Networks (Fourier-KAN), which leverage periodicity and strong nonlinearity to represent audio signals, eliminating the need for additional positional encoding. Furthermore, a Frequency-adaptive Learning Strategy (FaLS) is proposed to enhance the convergence of Fourier-KAN by capturing high-frequency components and preventing overfitting of low-frequency signals.  Extensive experiments conducted on natural speech and music datasets reveal that: (1) well-designed positional encoding and activation functions in Coordinate-MLPs can effectively improve audio representation quality; and (2) Fourier-ASR can robustly represent complex audio signals without extensive hyperparameter tuning. Looking ahead, the continuity and infinite resolution of implicit audio representations make our research highly promising for tasks such as audio compression, synthesis, and generation. The source code will be released publicly to ensure reproducibility. The code is available at \url{https://github.com/lif314/Fourier-ASR}.
\end{abstract}

\section{Introduction}
\begin{figure*}[!ht]
    \centering
    \includegraphics[width=0.99\textwidth]{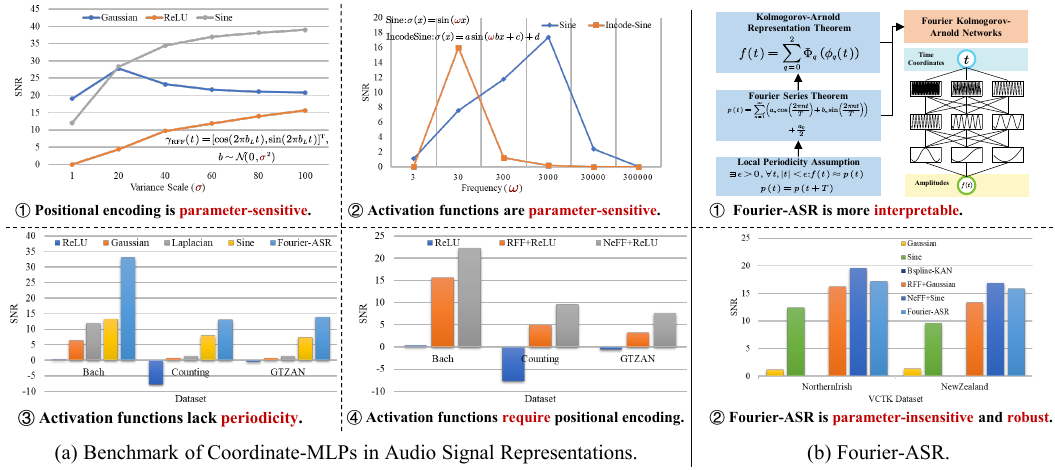}
    \caption{Properties of Coordinate-MLPs and Fourier-ASR. Validations are  in the appendix (\texttt{Appendix A}). }
    \label{fig:teaser}
\end{figure*}

Implicit Neural Representations (INRs) provide an innovative approach to signal parameterization by representing arbitrary discrete signals as continuous functions. These functions map the domain of the signal (coordinates, e.g., timestamps in audio) to the corresponding content at those coordinates (such as the amplitude of an audio signal). Typically, these functions are approximated using neural networks, and since current neural networks are primarily constructed using multilayer perceptrons (MLPs), these types of INRs are referred to as Coordinate-MLPs.

Compared to traditional discrete signal representation schemes, INRs offer continuous implicit representation that decouples from spatial resolution and allows for infinite resolution. Therefore, the storage required for parameterized signals is independent of spatial resolution, allowing these signals to be sampled at any desired resolution. Owing to these advantages, Coordinate-MLPs have been successfully applied to various modalities of data, including neural radiance fields \cite{mildenhall2020nerf}, 3D occupancy grids \cite{mescheder2019occupancynet}, Signed Distance Functions \cite{park2019deepsdf}, images \cite{sitzmann2020siren}, 2D computed tomography, and 3D magnetic resonance imaging \cite{tancik2020ffn, saragadam2023wire, kazerouni2023incode}.

Regarding audio signals, continuous representations offer the advantages of infinite resolution, enabling natural generation, efficient compression, and smooth processing. However, the representation of continuous audio signals using Coordinate-MLPs poses profound challenges due to the high noise, high frequency, nonlinearity, and local periodicity inherent in audio signals. According to the Weber-Fechner law, even relatively small reconstruction errors in audio signals can become perceptible due to the logarithmic nature of human auditory perception, thereby imposing high demands on the quality of audio reconstruction. Moreover, the simple combination of linear transformations and nonlinear activation functions in MLP networks makes it difficult to capture the periodicity and high-frequency components of audio signals. Through a comprehensive review, we find that till now only SIREN \cite{sitzmann2020siren} has attempted to represent audio signals using sinusoidal activation functions and provided a simple comparison with ReLU-MLPs, yet no further investigations have been conducted. 

To fill the gap, we establish, to our knowledge, the \textbf{first} open-source benchmarking framework to fully explore the potential and limitations of Coordinate-MLPs in continuous audio signal representations. Specifically, since the performance of a Coordinate-MLP is primarily determined by the choice of the activation function and optional positional encoding, we identify 3 types of positional encoding mappings and 16 commonly used activation functions from existing Coordinate-MLP methods. This results in 48 possible network configurations for audio signal representation, which we evaluate on speech and music datasets to assess their performance.
As shown in Fig. \ref{fig:teaser}(a), our benchmark reveals the following findings. (1) Most activation functions, except those with strong linearity (e.g., Gaussian) or periodicity (e.g., Sine), are unable to effectively represent audio signals. (2) Although some activation functions, such as Gaussian and Sine, are proposed to overcome spectral bias and the tedious parameter tuning associated with positional encoding, positional encoding remains indispensable for representing audio signals. It efficiently maps time coordinates to high-dimensional spaces, allowing the network to capture high-frequency components in audio signals. (3) Due to the local periodicity of audio signals, periodic activation functions (e.g., Sine) significantly outperform other activation functions in representational capacity. Moreover, incorporating Fourier feature-based positional encoding can further enhance their ability. (4) While Sine-type activation functions are effective at representing audio signals due to their periodic nature, they unfortunately require hyperparameter-sensitive positional encodings and frequency-dependent initialization schemes, which negatively impact their robustness and generalization capabilities.

The aforementioned issues of Coordinate-MLPs fundamentally arise from the inadequate nonlinearity and lack of periodicity inherent in MLPs. As illustrated in Fig. \ref{fig:teaser}(b), to enhance the nonlinear and periodic representational capabilities of neural networks, we propose a novel implicit audio representation framework, Fourier-ASR, based on the Fourier series theorem and the Kolmogorov-Arnold representation theorem. Firstly, we introduce a Kolmogorov-Arnold Network (Fourier-KAN) that utilizes Fourier basis functions to represent audio signals. This network implicitly decomposes any complex audio signal into a series of locally periodic Fourier series. Unlike MLPs, Fourier-KAN does not require additional positional encoding or activation functions, thereby avoiding cumbersome hyperparameter tuning. Furthermore, due to the use of Fourier basis functions, it more effectively captures the high-frequency components and local periodicity of signals. Secondly, to accelerate the convergence of Fourier-KAN, we introduce a Frequency Adaptive Learning Strategy (FaLS). FaLS employs an inverted frequency pyramid configuration to capture signals at various frequencies and utilizes a frequency-adaptive weight initialization scheme based on forward propagation theory to mitigate issues of gradient explosion or vanishing, thereby expediting convergence. Experimental results demonstrate that Fourier-ASR not only offers enhanced interpretability but is also robust to hyperparameter variations, effectively representing complex audio signals.

In summary, our contributions are summarized as follows:
\begin{itemize}
    \item We introduce the \textbf{first benchmark} for Coordinate-MLPs in audio representation, incorporating 3 types of positional encodings and 16 commonly used activation functions. Our benchmark provides an in-depth analysis of the impact of positional encoding and activation functions on the representation of continuous audio signals.
    
     \item To avoid spectral bias from positional encoding and complex parameter tuning of activation functions, we propose a novel audio signal representation framework, \textbf{Fourier-ASR}, based on the Fourier series theorem and the Kolmogorov-Arnold theorem. Fourier-ASR includes Fourier Kolmogorov-Arnold Networks (Fourier-KAN) and a Frequency-adaptive Learning Strategy (FaLS). Due to the periodicity and strong nonlinearity of Fourier basis functions, Fourier-ASR effectively represents audio signals and provides enhanced interpretability.
     
     \item As shown in Fig. \ref{fig:teaser}, extensive experiments conducted on speech and music datasets reveal that (1) careful tuning of positional encoding and activation function parameters can significantly enhance the representational capacity of Coordinate-MLPs for audio signals; and (2) Fourier-ASR can robustly represent audio signals without requiring cumbersome parameter tuning.
 \end{itemize}

\section{Related Work}
\begin{figure*}[!ht]
    \centering
    \includegraphics[width=0.97\textwidth]{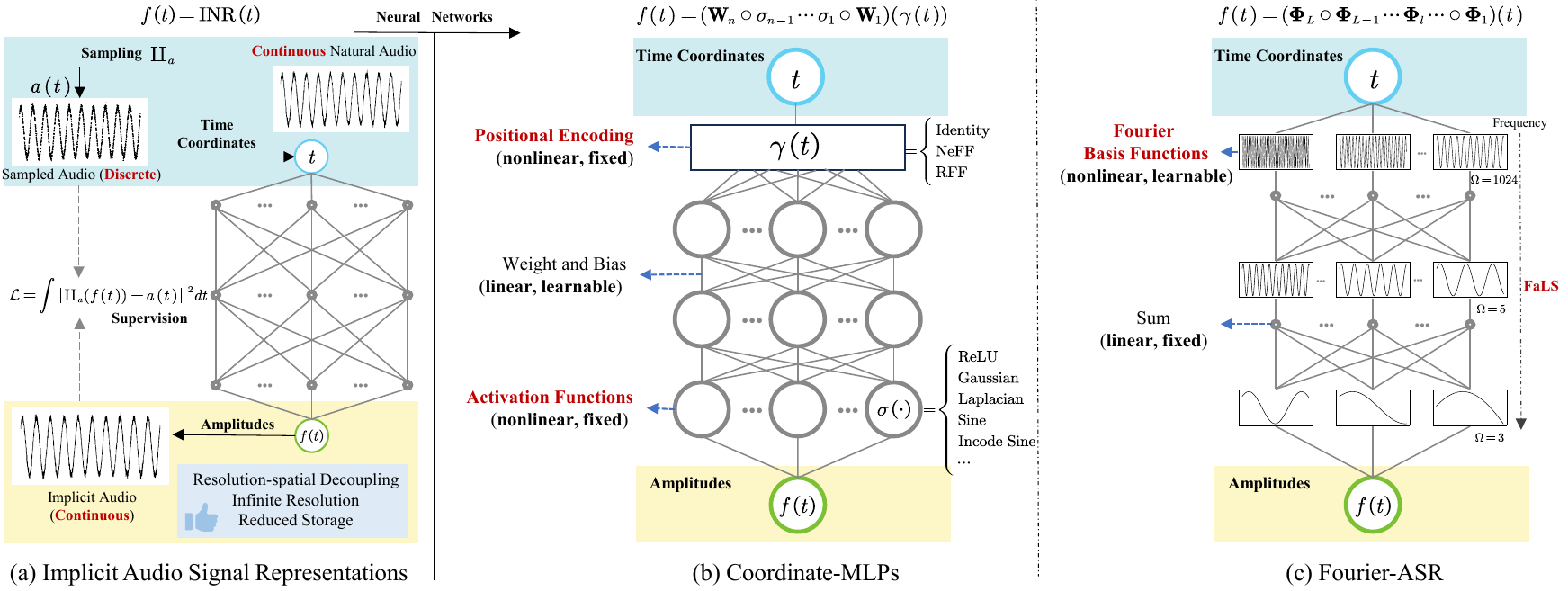}
    \caption{ (a) The problem definition of implicit audio representations; (b) The audio representation framework based on Coordinate-MLPs; (c) Fourier-ASR, a novel audio signal representation framework based on Fourier-KAN. }
    \label{fig:framework}
\end{figure*}

\subsubsection{Coordinate-MLPs.} The usage of Coordinate-MLPs differs significantly from that of traditional MLPs in two main aspects: (a) traditional MLPs typically operate on high-dimensional inputs, such as images, sounds, or 3D shapes; (b) traditional MLPs are primarily employed as classification heads, where the decision boundaries need not be smooth. In contrast, Coordinate-MLPs encode signals into weights, where the input is low-dimensional coordinates and the output must maintain smoothness. The success of NeRF \cite{mildenhall2020nerf} demonstrates that Coordinate-MLPs, when trained with a limited number of perspective images, can reconstruct photometric projections from any angle and at any resolution. This breakthrough spurs the application of Coordinate-MLPs in numerous fields, including radiance field reconstruction \cite{barron2022mipnerf360, Chen2022tensorf, muller2022ngp}, 3D shape representation \cite{wang2023neus, yu2022monosdf, yariv2023bakedsdf}, 2D image regression \cite{tancik2020ffn, saragadam2023wire, lindell2022bacon, ramasinghe2022activations}, audio signal regression \cite{sitzmann2020siren, kazerouni2023incode}, and inverse rendering problems in 2D CT and 3D MRI \cite{tancik2020ffn}.

\subsubsection{Positional Encoding.} 
Positional encoding facilitates the learning of high-frequency representations in radiance fields, images, and 3D shapes. NeRF \cite{mildenhall2020nerf} improves the ability of ReLU-MLPs to capture high-frequency signals by mapping the input coordinates to a high-dimensional Fourier space. Building upon NeRF, FFN \cite{tancik2020ffn} incorporates Gaussian noise to improve the robustness of ReLU-MLPs. Although positional encodings enable MLPs to represent high-frequency components, selecting the appropriate frequency scale is crucial and often involves cumbersome parameter tuning. Specifically, when the signal bandwidth is excessively increased, Coordinate-MLPs tend to produce noisy signal interpolations \cite{ramasinghe2022regularizingcoordinatemlps, hertz2021sape}.

\subsubsection{Activation Functions.}
The nonlinear representation capability of Coordinate-MLPs primarily arises from activation functions. In the field of INRs, various activation functions have been employed to approximate different types of signals. ReLU is frequently employed as the activation function in NeRF-related studies due to its simplicity and effective initialization scheme \cite{mildenhall2020nerf, barron2022mipnerf360, yu2021pixelnerf, Chen2022tensorf}. However, ReLU struggles to capture high-frequency information in radiance fields, necessitating additional positional encoding. To avoid the cumbersome parameter tuning associated with positional encoding, GARF \cite{chng2022garf} uses the Gaussian activation function, which can effectively capture high-frequency information but fails to capture periodic signals and tends to overfit both noise and signal equally. To address these issues, WIRE \cite{saragadam2023wire} utilizes the complex Gabor wavelet activation function to improve the robustness. SIREN \cite{sitzmann2020siren} employs the sine activation function to capture signal periodicity, though it is sensitive to initialization schemes, limiting its generalization to audio reconstruction tasks. Building on SIREN, INCODE \cite{kazerouni2023incode} makes the parameters of the sine activation functions learnable, thereby reducing the parameter sensitivity to some extent, but it still relies on the frequency-aware initialization scheme.

\section{Method}
\subsection{Problem Formulation}
As illustrated in  Fig. \ref{fig:framework}(a), natural audio signals are continuous functions of time, representing the variation in amplitude of sound signals over time. To convert this continuous signal into a digital format for storage and processing, the signal is discretely sampled, resulting in a discrete signal \( a(t) \) with respect to the time coordinate \( t \). However, in fields such as audio super-resolution, synthesis, and compression, researchers aim to leverage implicit neural representation techniques to preserve the continuity and differentiability of the signal as much as possible. Specifically, by receiving a discrete time coordinate \({t}\), a neural network regresses the amplitude \(f(t)\) corresponding to $t$, thereby encoding the audio signal within the network weights. We refer to this representation as the \textbf{Neural Amplitude Fields (NeAF)}. Optimization is performed by fitting \(f(t)\) to the sampled waveform \(a(t)\) using an MSE loss function,
\begin{equation}
    \mathcal{L}=\int \left\|\amalg_a(f({t}))-a({t})\right\|^2 d {t},
\end{equation} where $\amalg_a$ samples $f({t})$ at the waveform measurement locations. Given that NeAF is independent of spatial resolution, audio can be processed at any desired resolution.

\subsection{Benchmark of Coordinate-MLP-based NeAF}
As depicted in Fig. \ref{fig:framework}(b), to represent arbitrary complex audio signals, a $k$-layer Coordinate-MLP \( f: \mathbb{R} \rightarrow \mathbb{R} \) is employed, which takes the time coordinate \( t \in \mathbb{R}\) as input and outputs the amplitude \( f(t) \in \mathbb{R} \). Thus, $f(t)$ can be defined through the following recursive relations,
\begin{equation}
\begin{aligned}
\mathbf{z}^{(1)} & = \gamma({t}) \\
\mathbf{z}^{(i+1)} & =\sigma\left(\mathbf{W}^{(i)} \mathbf{z}^{(i)}+\mathbf{b}^{(i)}\right), i=1, \ldots, k-1 \\
f({t}) & =\mathbf{W}^{(k)} \mathbf{z}^{(k)}+\mathbf{b}^{(k)},
\end{aligned}
\end{equation} where \(\gamma(\cdot)\) denotes an optional positional encoding function that maps the input coordinate $t$ to a higher-dimensional space, \(\sigma(\cdot)\) represents the element-wise applied nonlinear activation function, \(\mathbf{W}^{(i)} \in \mathbb{R}^{d_{i+1} \times d_i}\) and \(\mathbf{b}^{(i)} \in \mathbb{R}^{d_{i+1}}\) denote the weights and biases of the \(i\)-th layer, respectively, while \(\mathbf{z}^{(i)} \in \mathbb{R}^{d_i}\) represents the hidden units of the \(i\)-th layer. 

Following the architecture of Coordinate-MLPs, we conduct an extensive review of implicit neural representations and identify 3 types of positional encoding and 16 potential activation functions. Specifically, as shown in Table \ref{tab:cmlps}, the three positional encoding schemes are identity mapping (\texttt{Identity}), NeRF Fourier features (\texttt{NeFF}) \cite{mildenhall2020nerf}, and random Fourier features (\texttt{RFF}) \cite{tancik2020ffn}. The primary activation functions include ReLU, Gaussian \cite{chng2022garf}, Laplacian \cite{ramasinghe2022activations}, Sine \cite{sitzmann2020siren}, Incode-Sine \cite{kazerouni2023incode}, and Gabor-Wavelet \cite{saragadam2023wire}, among others (details are provided in Table \ref{tab:mlp_benchmark}).

\begin{table}[!htbp]
    \centering
    \tabcolsep=0.0mm
    \fontsize{9}{11}\selectfont
    \begin{tabular}{ccc}
        \cline{1-3}
        PE ($\mathcal{P}$) & $\gamma \in \mathcal{P}$ & Parameter \\
        \cline{1-3}
        
        \texttt{Identity} & $\gamma({t})={t}$ & - \\

         \cdashline{1-3}
        \texttt{NeFF} & $\gamma({t})=[\cos (2^L \pi {t}), \sin (2^L \pi {t})]^{\mathrm{T}}$ & $[L]$ \\

         \cdashline{1-3}
        \multirow{2}{*}{\texttt{RFF}} & \multirow{2}{*}{\makecell{$\gamma({t})=[\cos (2 \pi {b}_L {t}), \sin (2 \pi {b}_L {t})]^{\mathrm{T}}, $ \\ ${b}_L \sim \mathcal{N}(0, \sigma^2)$}} & \multirow{2}{*}{\makecell{$[\sigma, L]$}} \\
        & \\
        
        \hline
        \hline
         Activations ($\mathcal{A}$) & $\sigma \in \mathcal{A}$ \\

         \hline
         \texttt{ReLU} & $\sigma(x)=\max(0, x)$ & - \\
         
         \cdashline{1-3}
         \texttt{Gaussian} & $\sigma(x)=e^{\frac{-x^2}{2a^2}}$ & $[a]$ \\

         \cdashline{1-3}
         \texttt{Laplacian} & $\sigma(x)=e^{\frac{-|x|}{a}}$  & $[a]$ \\

         \cdashline{1-3}
         \texttt{Sine} & $\sigma(x)=\sin(\omega x)$  &  $[\omega]$\\

         \cdashline{1-3}
         \texttt{Incode-Sine} & $\sigma(x)=a\sin(b\omega x + c) + d$ & $a,b,c,d, [\omega]$ \\

        \cdashline{1-3}
        $\cdots$ & $\cdots$ & $\cdots$  \\
         
         \cline{1-3}
    \end{tabular}
    \caption{The nonlinear mappings in Coordinate-MLPs. Note that \(a\) denotes a learnable parameter, while \([a]\) denotes a hyperparameter.  }
    \label{tab:cmlps}
\end{table}

It is noteworthy that Gaussian, Sine, and Incode-Sine activation functions are proposed to eliminate the dependence on positional encoding in radiance fields and image representations. However, high-dimensional positional encoding mappings may be beneficial for learning high-frequency features in audio signals and their structural variations at different time scales. Therefore, in our benchmark, we apply positional encoding mappings to all three activation functions. Based on Table \ref{tab:cmlps}, the Coordinate-MLPs used for benchmarking audio signal representations can be expressed as follows,
\begin{equation}
\label{eq:cmlp}
    f({t})=(\mathbf{W}_n \circ \sigma_{n-1} \cdots \sigma_1  \circ \mathbf{W}_1)(\gamma({t})), \\
    \gamma(\cdot) \in \mathcal{P}, \sigma_i(\cdot) \in \mathcal{A},
\end{equation} where \({t}\) denotes the input time coordinate normalized to the interval $[0,1]$, \(\mathcal{P}\) represents the set of positional encodings, and \(\mathcal{A}\) denotes the set of activation functions. 

\subsection{Fourier-ASR}
Our benchmark indicates that only through carefully designed positional encoding and activation functions can some Coordinate-MLPs effectively represent audio signals. However, their flexibility and generality are reduced due to complex parameter tuning and high sensitivity to initialization. To address this issue, as shown in Fig. \ref{fig:framework}(c), drawing from the Fourier series theorem and the  Kolmogorov-Arnold representation theorem, we introduce a novel framework for audio signal representation, Fourier-ASR, which incorporates Fourier Kolmogorov-Arnold Networks (Fourier-KAN) and a Frequency-adaptive Learning Strategy (FaLS).

\subsubsection{Fourier Kolmogorov-Arnold Networks (Fourier-KAN).}
Unlike Coordinate-MLPs based on the Universal Approximation Theorem \cite{HORNIK1989359}, which use combinations of linear transformations and nonlinearities, Fourier-ASR follows the Kolmogorov-Arnold Representation Theorem \cite{Kolmogorov1956, Arnold1957} to represent any continuous function as a finite composition of single-variable functions and addition. For a continuous signal \( f(t) \), this simplifies to,
\begin{equation}
    f({t}) = \sum_{q=0}^{2} \Phi_q \left(\phi_{q}(t) \right),
\end{equation}
where \(\Phi_q: \mathbb{R} \rightarrow \mathbb{R}\) and \(\phi_{q}: [0,1] \rightarrow \mathbb{R}\) denote the outer and inner functions, respectively.
To enhance the capacity and learnability of this representation, we employ the KAN \cite{liu2024kan} approach to extend the network to an arbitrary number of layers.

\begin{table}[!ht]
    \centering
    \resizebox{0.48\textwidth}{!}{
    \begin{tabular}{|c|}
        \hline
        \begin{minipage}[t]{\linewidth}
            \begin{assumption}\label{asp:lpa}
                \textbf{Local Periodicity Assumption.} 
                For a complex non-stationary signal \( f(t) \), there exists a sufficiently small time interval \( \epsilon > 0 \) where \( f(t) \) can be approximated by a periodic function \( p(t) \):
                $$
                \exists \epsilon > 0, \quad \forall t, \quad |t| < \epsilon: \quad f(t) \approx p(t)
                $$
                where \( p(t) \) is a periodic function with period \( T \), satisfying \( p(t + T) = p(t) \).
            \end{assumption}
        \end{minipage} \\ 
        \hdashline
        \begin{minipage}[t]{\linewidth}
            \begin{theorem}\label{thm1}
               \textbf{Fourier Series Theorem.} 
                Any periodic signal \( p(t) \) with period \( T \) can be represented as an infinite series of sine and cosine functions,
                $$
                p(t) = \frac{a_0}{2} + \sum_{n=1}^{\infty} \left( a_n \cos\left(\frac{2\pi nt}{T}\right) + b_n \sin\left(\frac{2\pi nt}{T}\right) \right)
                $$
                where \( a_0 \), \( a_n \), and \( b_n \) are the Fourier coefficients.
            \end{theorem}
        \end{minipage} \\ 
        \hline
    \end{tabular}
    }
\end{table}

Specifically, consider a neural network with a shape of $\left[n_0, n_1, \cdots, n_L\right]$, where $n_l$ denotes the number of neurons on the $l$-th layer of the computational graph. For the $i$-th node on the $l$-th layer, denoted by $(l, i)$, the activation value of this neuron is $t_{l,i}$. Between the $l$-th and $(l+1)$-th layers, there are $n_l \times n_{l+1}$ non-linear basis functions. Based on Assumption \ref{asp:lpa} and Theorem \ref{thm1}, any audio signal within short time intervals can be approximated as combinations of cosine and sine functions. Therefore, unlike using B-Splines in KAN \cite{liu2024kan}, we employ Fourier basis functions as the non-linear units connecting neurons $(l, i)$ and $(l+1, j)$,
\begin{equation}
    \label{eq:fourier_basis}
    \begin{split}
        \phi_{l, j, i} (t_{l,i}) = a_{l, i} \cos(\omega t_{l, i}) + b_{l, i} \sin(\omega t_{l, i}) + c_{l, i}, \\
        l=0, \cdots, L-1, i=1, \cdots, n_l,  j=1, \cdots, n_{l+1},
    \end{split}
\end{equation} where \(a_{l,i}, b_{l,i}\) are learnable Fourier coefficients, \(c_{l,i}\) is a learnable bias term, and \(\omega\) is a frequency hyperparameter.
Then, the activation value $t_{l+1, j}$ of the $(l+1, j)$ neuron is simply the sum of all incoming post-activations,
\begin{equation}
\label{eq:post_neuron}
t_{l+1, j}=\sum_{i=1}^{n_l} \phi_{l, j, i}\left(t_{l, i}\right), j=1, \cdots, n_{l+1} .
\end{equation}
For the $l$-th Fourier KAN layer, by rewriting Eq. \ref{eq:post_neuron} under the matrix form, we can have,  
$$
\mathbf{t}_{l+1}=\underbrace{\left(\begin{array}{cccc}
\phi_{l, 1,1}(\cdot) & \phi_{l, 1,2}(\cdot) & \cdots & \phi_{l, 1, n_l}(\cdot) \\
\phi_{l, 2,1}(\cdot) & \phi_{l, 2,2}(\cdot) & \cdots & \phi_{l, 2, n_l}(\cdot) \\
\vdots & \vdots & & \vdots \\
\phi_{l, n_{l+1}, 1}(\cdot) & \phi_{l, n_{l+1}, 2}(\cdot) & \cdots & \phi_{l, n_{l+1}, n_l}(\cdot)
\end{array}\right)}_{\boldsymbol{\Phi}_l} \mathbf{t}_l.
$$  where \(\mathbf{\Phi}_{l}\) is the transition matrix between the Fourier layers.

In summary, Fourier-ASR employs Fourier-KAN to derive continuous representations from discrete audio signals as,
\begin{equation}
    f({t})=(\mathbf{\Phi}_L \circ \mathbf{\Phi}_{L-1} \cdots \mathbf{\Phi}_{l}  \cdots \circ \mathbf{\Phi}_{1})({t}).
\end{equation} Compared to the Coordinate-MLPs (Eq. \ref{eq:cmlp}), our Fourier-KAN leverages Fourier basis functions to achieve not only enhanced nonlinear representation capabilities but also the ability to capture local periodicity in audio signals.

\subsubsection{Frequency-adaptive Learning Strategy (FaLS).}
Due to the varying frequency distributions of audio signals across different time scales, a fixed frequency hyperparameter (\(\omega\) in Eq. \ref{eq:fourier_basis}) can lead Fourier-KAN to predominantly learn specific frequency components, thereby hindering convergence. To address this issue, we propose a Frequency-adaptive Learning Strategy (FaLS). Specifically, we assign basis functions with varying frequency thresholds to different Fourier-KAN layers. Then, a Fourier-KAN can be represented as,
\begin{equation*}
\begin{aligned}
    \mathbf{z}^{(0)} & = {t} \\
    \mathbf{z}^{(l+1)} &= \sum_{\omega=1}^{\Omega_l} \left[ \mathbf{a}^{(l,\omega)}\cos(\omega \mathbf{z}^{(l)}) + \mathbf{b}^{(l,\omega)}\sin(\omega \mathbf{z}^{(l)})\right] + \mathbf{c}^{(l)} \\
    f({t}) & = \mathbf{z}^{(n_L)}\left(\mathbf{z}^{(n_L-1)}\left(\mathbf{z}^{(l)}\left(...(\mathbf{z}^{(0)})\right)\right)\right), l=n_0, \ldots, n_L, \\
\end{aligned}
\end{equation*} where \(\mathbf{a}^{(l, \omega)}\) and \(\mathbf{b}^{(l, \omega)} \in \mathbb{R}^{d_{l+1} \times d_l}\) denote the Fourier coefficient weights for the \(l\)-th layer at frequency \(\omega\), \(\mathbf{c}^{(l)} \in \mathbb{R}^{d_{l+1}}\) is the bias term of the $l$-th layer, \(z^{(l)} \in \mathbb{R}^{d_l}\) denotes the hidden units of the \(l\)-th layer, and \(\Omega_l\) is a hyperparameter indicating the maximum frequency threshold for the \(l\)-th layer.

\textbf{\textit{Parameter initialization.}} Following the principles of Xavier \cite{pmlr2010xavierinit} and Kaiming's work \cite{he2015kaiminginit}, we derive the initialization scheme for the Fourier-KAN. Specifically, in the forward propagation process at layer \( l \) of the Fourier-KAN, the symmetry of the Fourier basis functions ensures that the expected values of both the input and the output are zeros, i.e., \( E[\mathbf{z}^{(l)}] = E[\mathbf{z}^{(l+1)}] = 0 \). According to Kaiming initialization \cite{he2015kaiminginit}, we make the following assumptions: (1) the expected values of the Fourier parameters \(\mathbf{a}^{(l, \omega)}\) and \(\mathbf{b}^{(l, \omega)}\) are both zeros, and the bias term \(\mathbf{c}^l\) is omitted; (2) the variances of the input \(\mathbf{z}^{(l)}\)  and the output \(\mathbf{z}^{(l+1)}\) are both ones. Thereby, we can determine the variance of the output at layer \( l \) as,
\begin{equation*}
    \begin{aligned}
    Var&[\mathbf{z}^{(l+1)}] = Var[\sum_{\omega=1}^{\Omega_l} \left[ \mathbf{a}^{(l,\omega)}\cos(\omega \mathbf{z}^{(l)}) + \mathbf{b}^{(l,\omega)}\sin(\omega \mathbf{z}^{(l)})\right]] \\
    &=\sum_{\omega=1}^{\Omega_l} \left( \cos^2(\omega \mathbf{z}^{(l)})Var[\mathbf{a}^{(l, \omega)}] + \sin^2(\omega \mathbf{z}^{(l)}) Var[\mathbf{b}^{(l, \omega)}] \right). \\
    \end{aligned}
\end{equation*} Assuming that the variances of the Fourier coefficients are equal, we can have,
\begin{equation}
    Var[\mathbf{a}^{(l)}]=Var[\mathbf{b}^{(l)}]=\frac{1}{\Omega_l}. 
\end{equation}
Thus, each independent Fourier coefficient \(a_i^{(l)}\) (and \(b_i^{(l)}\)) is initialized using the following normal distribution,
\begin{equation}
    a_i^{(l)}, b_i^{(l)} \sim \mathcal{N}(0, \frac{1}{\Omega_l d_{in}^{(l)}}), 
\end{equation} where \(d_{in}^{(l)}\) denotes the dimensionality of the input to layer \( l \).

\begin{table*}[!ht]
\centering
\tabcolsep=1.3mm
\fontsize{9}{11}\selectfont
    \begin{tabular}{c|c|c|c|c|c|c|c|c|c|c|c}
  
        \cline{1-12}
        \multirow{2}{*}{\makecell{Activation $\sigma(\cdot)$}} & \multirow{2}{*}{\makecell{Equation}} & \multirow{2}{*}{\makecell{Parameter}} & \multirow{2}{*}{\makecell{PE $\gamma(\cdot)$}} & \multicolumn{2}{c|}{Bach \texttt{(7s)}} & \multicolumn{2}{c|}{Counting \texttt{(7s)}} & \multicolumn{2}{c|}{Blues \texttt{(30s)}} & \multicolumn{2}{c}{\texttt{Avg.}} \\
        
        \cdashline{5-12}
        & & & & SNR $\uparrow$ & LSD $\downarrow$ & SNR $\uparrow$ & LSD $\downarrow$ & SNR $\uparrow$ & LSD $\downarrow$ & SNR $\uparrow$ & LSD $\downarrow$ \\
        \cline{1-12}
        \multirow{3}{*}{\makecell{$\operatorname{PReLU}$}} & \multirow{3}{*}{\makecell{\begin{tabular}{l} 
        $\begin{cases}x, & \text{ if } x>0 \\ ax, & \text{otherwise} \end{cases}$
        \end{tabular}}} & \multirow{3}{*}{\makecell{$[a]$ }}& \texttt{Identity} & 0.00 & 4.724 & 0.00 & 4.630 & 0.00 & 7.031 & 0.00 & 5.462 \\
        & & & \texttt{RFF} & 13.42 & 1.010 & 3.38 & 1.437 & 2.50 & 2.035 & 6.43 & 1.494  \\
        & & & \texttt{NeFF} & 17.50 & 1.133 & 7.88 & 1.575 & 5.20 & 1.539 & 10.19 & 1.416 \\
        
        \cline{1-12}
        \multirow{3}{*}{\makecell{$\operatorname{ReLU}$}} & \multirow{3}{*}{\makecell{$\max(0, x)$}} & & \texttt{Identity} & 0.00 & 4.623 & -7.66 & 4.546 & 0.00 & 6.774 & -2.55 & 5.314 \\
        & & &  \texttt{RFF} & 15.62 &  0.978 & 4.93 & \fs 1.400 & 3.23 & 1.862 & 7.93 &  1.413 \\
        & & &  \texttt{NeFF} &  22.29 & 1.129 &  9.57 &  1.538 & 7.64 & 1.324 &  13.17 &  1.330 \\
        
        \cline{1-12}
        \multirow{3}{*}{\makecell{$\operatorname{Gaussian}$}} & \multirow{3}{*}{\makecell{$e^{\frac{-x^2}{2a^2}}$ }} & \multirow{3}{*}{ \makecell{$[a]$ }} & \texttt{Identity} & 6.35 & 1.130 & 0.74 & 2.165 &  0.68 &  3.059 & 2.59 & 2.118 \\
        & & & \texttt{RFF} & 20.85 & 2.046 & 12.14 & 3.195 &  11.80 & 1.346 & 14.93 & 2.196 \\
        & & & \texttt{NeFF} & 19.68 & 2.127 & 9.20 & 3.438 & 7.74 & 1.597 & 12.21 & 2.387 \\
        
       \hline
        \multirow{3}{*}{\makecell{$\operatorname{Laplacian}$}} & \multirow{3}{*}{\makecell{
        $e^{\frac{-|x|}{a}}$}} & \multirow{3}{*}{\makecell{$[a]$ }} & \texttt{Identity} &  12.04 &  0.932 &  1.34  & \fs 1.561 & \underline{ 1.37} & \underline{ 2.434} &  4.92 &  \underline{1.642} \\
        & & & \texttt{RFF} & 15.57 & 2.386 & 10.97 & 2.632 &  14.74 & \fs 1.112 & 13.76 & 2.043 \\
        & & & \texttt{NeFF} & 15.26 & 2.434 & 8.67 & 3.191 &  8.16 &  1.526 & 10.70 & 2.384 \\
        
        \hline
        \multirow{3}{*}{\makecell{Sine}} & \multirow{3}{*}{\makecell{
        $\sin (\omega x)$}} & \multirow{3}{*}{\makecell{$[\omega]$}}& \texttt{Identity} & \underline{ 13.36} & \underline{ 0.838} & \underline{ 7.96} &  1.660 & \fs 7.47 & \fs 1.722 & \fs 9.59 & \fs 1.407 \\
        & & & \texttt{RFF} & \fs 39.02 & \fs 0.582 & \fs 13.06 & \underline{ 1.412} & \fs 16.57 & \underline{ 1.156} & \fs 22.88 & \fs 1.050 \\
        & & & \texttt{NeFF} & \fs 42.39 & \fs 0.537 & \fs 33.58 & \fs 0.914 & \fs 22.02 & \fs 0.696 & \fs 32.66 &  \fs 0.716 \\
        
        \hline
        \multirow{3}{*}{\makecell{Incode-Sine}} & \multirow{3}{*}{\makecell{
        $a\sin (b \omega x + c) + d$}} & \multirow{3}{*}{\makecell{$[\omega], a,b,c,d$ }}& \texttt{Identity} & \fs 15.98 & \fs 0.778 & \fs 8.16 & \underline{ 1.611} & 0.01 & 3.865 & \underline{ 8.05} &   2.085 \\
        & & & \texttt{RFF} & \underline{ 38.10} & \underline{ 0.595} & \underline{ 12.86} &  1.559 & \underline{ 15.13} &  1.241 & \underline{ 22.03} &  \underline{ 1.132} \\
        & & & \texttt{NeFF} & \underline{ 41.40} & \underline{ 0.556} & \underline{ 32.24} & \underline{ 1.038} & \underline{ 21.33} & \underline{ 0.763} & \underline{ 31.99} & \underline{ 0.786} \\ 
         \cline{1-12}
    \end{tabular}
    \caption{Benchmark leaderboard of Coordinate-MLPs. For different positional encodings (\texttt{Identity}, \texttt{RFF}, \texttt{NeFF}), the best results are bold for first and underlined for second. Note that ``\(a\)'' denotes a learnable parameter, while ``\([a]\)'' denotes a hyperparameter. The benchmarking results for the remaining 10 activation functions are provided in the appendix (\texttt{Appendix D}). }
    \label{tab:mlp_benchmark}
\end{table*}

\textbf{\textit{Inverted pyramid frequency setting.}}
Given the depth \(L\) and width of a Fourier-KAN, the hyperparameters \([\Omega_0, \cdots, \Omega_l, \Omega_L]\) dictate the number of Fourier basis functions and the tendency to learn frequency components in each layer. With the same network capacity, a larger \(\Omega\) enhances the frequency resolution, improving the network's ability to capture audio signal periodicity and fluctuations. Similar to the role of positional encoding in Coordinate-MLPs, an inverted pyramid frequency setting is beneficial for the Fourier-KAN in capturing high-frequency information, thereby accelerating convergence. For instance, a 3-layer Fourier-KAN with \(\Omega\) set to $[64, 5, 3]$ outperforms \([8, 8, 8]\), which may lead to convergence issues.

\section{Experiments}
\subsection{Experimental Setup}
\subsubsection{Datasets.} 
\textbf{GTZAN} music dataset \cite{tzanetakis2001gtzan} includes 1000 thirty-second music clips across ten genres. \textbf{CSTR VCTK} speech corpus \cite{yamagishi2019cstr} consists of voice recordings from 110 speakers with diverse accents, each speaking approximately 400 sentences. For the benchmark, we used two 7-second clips provided by SIREN \cite{sitzmann2020siren} (``Bach" and ``Counting") and a 30-second clip from GTZAN 
 dataset (``Blues").  To comprehensively evaluate the performance of effective methods, we selected ten audio clips of different genres from the GTZAN dataset and ten audio clips with various accents from the CSTR VCTK dataset.

\subsubsection{Networks.} We ensured that the network parameters were comparable, ranging between 250K and 270K. For the Coordinate-MLPs, each network has a depth of 6 and a width of 256. In contrast, the Fourier-ASR network has a depth of 6 and a width of 64, with the maximum frequency thresholds set to 1024, 5, and 3 for the input layer, hidden layers, and the output layer, respectively.

\begin{table*}[!htbp]
  \tabcolsep=1.4mm
\fontsize{9}{11}\selectfont
  \begin{tabular}{c|ccccccccccccc}
    \cline{1-14}
    \multicolumn{2}{c}{\textbf{GTZAN Dataset}} & Metrics & \small \texttt{blu.} & \texttt{cla.}  &  \texttt{cou.} & \texttt{dis.} & \texttt{hip.} & \texttt{jaz.} & \texttt{met.} & \texttt{pop.} & \texttt{reg.} & \texttt{roc.} & \texttt{Avg.} \\
   \cline{1-14}
    \multirow{6}{*}{Baselines} & \multirow{2}{*}{\makecell{Gaussian (MLP)}} & SNR $\uparrow$ & 0.68 & 0.25 & 2.64 & 2.40 & 1.02 & 0.15 & 1.66 & 4.61 & 1.33 & 1.06 & 1.58  \\
    &  & LSD $\downarrow$ & 3.059 & 3.175 & 3.028 & 3.379 & 3.761 & 3.494 &3.634 & 3.030 & 3.047 & 3.219 & 3.383 \\
   \cdashline{2-14}

   & \multirow{2}{*}{\makecell{Sine (MLP)}} & SNR $\uparrow$ & 7.47 & 2.86 & 5.92 & 7.34 & 3.04 & 6.08 & 4.32 & 8.776 & 4.87 & 6.84 & 5.76  \\
    &  & LSD $\downarrow$ & 1.722 & 1.755 & 2.338 & 2.204 & 2.754 & 1.771 & 2.830 & 2.595 & 1.969 & 1.900 & 2.184 \\
   \cdashline{2-14}

    & \multirow{2}{*}{\makecell{B-Spline (KAN)}} & SNR $\uparrow$ & 0.00 & 0.01 & 0.00 & 0.00  & 0.17 & 0.00& 0.00  & 2.07 & 0.00 & 0.01 & 0.23  \\
    &  & LSD $\downarrow$ & 4.643 & 4.278 & 6.685 & 5.799 & 4.359 & 4.899  & 6.000 & 3.864 & 7.533 & 4.899 & 5.30 \\
   \hline
   
    \multirow{6}{*}{Ours}  & \multirow{2}{*}{\makecell{RFF+Gaussian (MLP)}} & SNR $\uparrow$ & 11.80& 10.76 & 11.98 & 12.00 & 11.30 & 12.75 & 11.25 & 12.07 & 11.57 & 11.84 & 11.73 \\
    &  & LSD $\downarrow$ & 1.346 & 1.721 & 1.474 & 1.299 &  1.148 & 1.936 & 1.731 & 1.439 & 1.362 & 1.481 & 1.494  \\
   \cdashline{2-14}
    
    & \multirow{2}{*}{\makecell{NeFF+Sine (MLP)}} & SNR $\uparrow$ & \fs 22.02 & \fs 25.95 & \fs 16.35 & \fs 17.70 & \fs 13.92 & \fs 19.22 & \fs 13.05 & \fs 15.27 & \fs 17.79 & \fs 19.16 & \fs 18.04 \\
    &  & LSD $\downarrow$ & \fs 0.696 &\fs  0.585 & \fs 1.064 & \fs 1.036 & \fs 0.741 & \fs 0.983 & \fs 0.902 &\fs  1.245 & \fs 0.883 & \fs 0.714 & \fs 0.885 \\
   \cdashline{2-14}

    & \multirow{2}{*}{\makecell{Fourier-ASR (KAN)}} & SNR $\uparrow$ &\underline{13.80} &\underline{ 15.05} &\underline{ 12.54} &\underline{ 12.87} &\underline{ 12.22} &\underline{ 13.67} &\underline{  12.21} &\underline{ 13.27} & \underline{ 12.42} &\underline{ 12.65} & \underline{ 12.76}  \\
    &  & LSD $\downarrow$ &\underline{ 1.245} &\underline{ 0.913} &\underline{ 1.249} &\underline{ 1.158} &\underline{ 1.059} &\underline{ 1.244} &\underline{ 1.203} & \underline{ 1.302} &\underline{ 1.110} &\underline{ 1.399} & \underline{ 1.110} \\
 
    \hline
    \hline

    \multicolumn{2}{c}{\textbf{CSTR VCTK Dataset}} & Metrics & \texttt{p225} & \texttt{p234}  &  \texttt{p238} & \texttt{p245} & \texttt{p248} & \texttt{p253} & \texttt{p335} & \texttt{p345}  & \texttt{p363} & \texttt{p374}  & \texttt{Avg.} \\
   \cline{1-14}

    \multirow{6}{*}{Baselines} & \multirow{2}{*}{\makecell{Gaussian (MLP)}} & SNR $\uparrow$ & 1.88 & 2.09  & 1.16 & 4.06 & 0.23 & 2.39 & 1.37 & 3.06 & 5.56 & 2.25 & 2.41 \\
    &  & LSD $\downarrow$ & 2.126 & 2.034 & 2.557 & 1.831 & 2.884 & 2.065 & 2.277 & 1.896 & 1.791 & 1.827 & 2.129  \\
   \cdashline{2-14}

   & \multirow{2}{*}{\makecell{Sine (MLP)}} & SNR $\uparrow$ & 14.86 & 10.88 & 12.38 & 14.41 & 10.32 & 13.85 & 9.61 & 15.89 & 12.78 & 12.53 & 12.75  \\
    &  & LSD $\downarrow$ & 1.743 & 1.588 & 1.748 & 1.665 & 1.630 & 1.672 & 1.619 &  1.556 & 1.716 & 1.500 & 1.644 \\
   \cdashline{2-14}

    & \multirow{2}{*}{\makecell{B-Spline (KAN)}} & SNR $\uparrow$ & 0.01 & 0.02 & 0.01 & 0.01  & 0.00  & 0.01 & 0.02 & 0.02 & 0.05 & 0.11 & 0.03  \\
    &  & LSD $\downarrow$ & 3.312 & 3.113 & 3.317 & 3.151 & 3.506 & 3.160 & 3.000 & 2.957 & 2.705 & 2.631 & 3.085 \\
   \hline
   
   \multirow{6}{*}{Ours}  & \multirow{2}{*}{\makecell{RFF+Gaussian (MLP)}} & SNR $\uparrow$ & 11.67 & 12.93 & 16.19 & 11.99 & 15.52 & 12.21 & 13.32 & 15.95 & 12.79 & 12.28 & 12.81  \\
    &  & LSD $\downarrow$ & 2.401 & 2.128 & 1.789 & 2.218 & 2.183 & 2.258 & 2.076 & 1.704 & 2.012 & 2.059 & 1.983 \\
   \cdashline{2-14}
    
    & \multirow{2}{*}{\makecell{NeFF+Sine (MLP)}} & SNR $\uparrow$ & \fs 25.20 & \fs 31.63 & \fs 19.56 & \fs 32.03  & \fs 27.00 & \fs 27.11 & \fs 16.87 & \fs 28.38 & \fs 29.25 & \fs 30.83 & \fs 26.79   \\
    &  & LSD $\downarrow$ & \fs 1.015 &\fs 0.734 & \fs 1.235 & \fs 0.866 &\fs 1.134 &\fs 0.917 &\fs 1.207 &\fs 1.032  &\fs 0.753 &\fs 0.877 &\fs 0.977 \\
   \cdashline{2-14}

   & \multirow{2}{*}{\makecell{Fourier-ASR (KAN)}} & SNR $\uparrow$ & \underline{ 18.30}  & \underline{ 20.68} &\underline{ 17.12} &\underline{ 18.26} &\underline{ 21.34} &\underline{ 17.40} &\underline{ 15.79} &\underline{ 17.34} &\underline{ 17.86} &\underline{ 20.20} &\underline{ 18.43} \\
   &  & LSD $\downarrow$ &\underline{ 1.495} &\underline{ 1.228} &\underline{ 1.615} &\underline{ 1.310} &\underline{ 1.464} &\underline{ 1.397} &\underline{ 1.456 }&\underline{ 1.417} &\underline{ 1.321} &\underline{ 1.267} &\underline{ 1.397} \\

    \hline
   
  \end{tabular}
  \caption{Evaluation of Fourier-ASR and new nonlinear mapping designs on GTZAN and CSTR VCTK dataset.}
  \label{tab:audio}
\end{table*}

\subsubsection{Evaluation Metrics.} 
Signal-to-Noise Ratio (SNR) \cite{roux2018snr} and Log-Spectral Distance (LSD) \cite{gray1976lsd} were utilized to assess the temporal and spectral errors in the reconstructed audios, respectively. Since LSD provides an indirect measure for frequency domain evaluation, we primarily focus on the SNR metric.

\subsection{Benchmark Leaderboard}
We begin by examining the impact of nonlinear mappings, which are commonly presumed but have not yet been analyzed in the context of implicit audio representation. In line with Eq. \ref{eq:cmlp}, Table \ref{tab:mlp_benchmark} presents the evaluation results of audio signal representation using 16 different activation functions and 3 types of positional encoding (\texttt{Identity}, \texttt{NeFF}, and \texttt{RFF}). Based on this comprehensive benchmarking, the following conclusions can be drawn:
\begin{itemize}
    \item Most activation functions (Sigmoid, ReLU, Tanh, etc.), aside from those with strong nonlinearity (Gaussian-type) and periodicity (Sine-type), fail to capture the high-frequency and local periodicity of audio signals.
    \item Positional encodings significantly enhance the ability of Coordinate-MLPs to represent audio signals due to their high-dimensional mapping capabilities, which improve the model's ability to capture high-frequency information. This enhancement is particularly notable for Gaussian (11.02dB \textcolor{red}{$\uparrow$} in SNR) and Sine (18.96dB \textcolor{red}{$\uparrow$} in SNR) activation functions.
    \item In the context of positional encoding, the introduction of random Gaussian noise by \texttt{RFF} makes it more suited to Gaussian-type activation functions ($\approx 3$dB \textcolor{red}{$\uparrow$} in SNR). Conversely, \texttt{NeFF} employs Fourier mappings, which are more compatible with Sine-type activation functions ($\approx9$dB \textcolor{red}{$\uparrow$} in SNR).
\end{itemize}

\subsection{Evaluation of Fourier-ASR and New Designs}
Based on the benchmark leaderboard presented in Table \ref{tab:mlp_benchmark}, we selected effective nonlinear mappings for comparison with Fourier-ASR on the GTZAN \cite{tzanetakis2001gtzan} and CSTR VCTK \cite{yamagishi2019cstr} datasets. It is noteworthy that although Gaussian \cite{ramasinghe2022activations} and Sine \cite{sitzmann2020siren} activation functions were introduced to mitigate the complex parameter adjustments and pectral bias associated with positional encoding, we found that positional encoding remains essential due to the high-frequency nature and local periodicity of audio signals. Consequently, we designed \textbf{new nonlinear mappings}, namely \texttt{RFF+Gaussian} and \texttt{NeFF+Sine}, to address these challenges.

As shown in Table \ref{tab:audio}, the designs \texttt{RFF+Gaussian} and \texttt{NeFF+Sine} significantly enhance the ability of Coordinate-MLPs to represent audio signals. On the GTZAN dataset, these methods improve the SNR by 10.15dB \textcolor{red}{$\uparrow$} and 12.28dB \textcolor{red}{$\uparrow$}, respectively. On the CSTR VCTK dataset, the SNR improvements are 10.40dB \textcolor{red}{$\uparrow$} and 14.04dB \textcolor{red}{$\uparrow$}, respectively. Due to the periodic nature of Fourier basis functions and the Frequency-adaptive Learning Strategy (FaLS), our proposed Fourier-ASR(KAN) significantly outperforms Sine(MLP) ($\approx6$dB \textcolor{red}{$\uparrow$}) and B-Spline(KAN) ($\approx18$dB \textcolor{red}{$\uparrow$}). However, because existing optimization strategies are not perfectly adapted to KAN networks \cite{liu2024kan}, Fourier-ASR(KAN) performs slightly worse than the locally periodic NeFF+Sine(MLP). Nonetheless, Fourier-ASR(KAN) does not require positional encoding, thereby avoiding the need for cumbersome hyperparameter tuning.

\section{Conclusion and Future Work}
We proposed the first open-source benchmark for evaluating implicit neural audio signal representations based on Coordinate-MLPs, addressing a critical gap in standardized performance assessment. We demonstrated the effectiveness of combining positional encoding and nonlinear mapping designs of activation functions in the field of continuous audio representations. Additionally, we introduced a novel audio signal representation framework, Fourier-ASR, which integrates the Fourier series theorem and the Kolmogorov-Arnold representation theorem, offering enhanced interpretability and more stable representational capacity. Our work not only guides the selection of components for Coordinate-MLP-based audio signal representations but also advances the development of audio representation applications. Due to the superior characteristics of implicit neural representations, such as continuous differentiability and decoupling from spatial resolution, our work can be effectively applied to downstream tasks such as audio super-resolution, denoising, compression, and generation.

\section*{Acknowledgments}
This work was supported in part by the National Natural Science Foundation of China under Grant 62272343; in part by the Shuguang Program of Shanghai Education Development Foundation and Shanghai Municipal Education Commission under Grant 21SG23; and in part by the Fundamental Research Funds for the Central Universities.

\bibliography{aaai25}


\newpage
\clearpage
\begin{center}
\textbf{Contents}
\end{center}

A. Properties of Coordinate-MLPs 

    \qquad A.1 Positional encoding is parameter-sensitive
    
    \qquad A.2 Activation functions are parameter-sensitive
    
    \qquad A.3 Periodic activation functions are sensitive to
    
    \qquad \quad initialization schemes

B. Memorization
    
    \qquad B.1 Memorization of Coordinate-MLPs
    
    \qquad B.2 Memorization of Fourier-KAN
    
    \qquad B.3 Parameter Utilization Comparison

C. Interpretability

    \qquad C.1 Interpretability of Coordinate-MLPs
    
    \qquad C.2 Interpretability of Fourier-KAN

D. Further Experiments

    \qquad D.1 Experimental Setup

    \qquad D.2 Supplement to Benchmark Leaderboard

    \qquad D.3 Qualitative Experiments
    
References

\section{A. Properties of Coordinate-MLPs}

\subsection{A.1 Positional encoding is parameter-sensitive}
\textbf{Analysis.}
Research on implicit neural representations has demonstrated that networks employing activation functions such as ReLU exhibit spectral bias \cite{pmlrv97rahaman19a}. These networks tend to first learn low-frequency functions, as these functions exhibit global behavior and are more stable for optimization. However, spectral bias also hinders the network's ability to learn high-frequency information. To address this issue, NeRF \cite{mildenhall2020nerf} proposes mapping input coordinates into a higher-dimensional space to enable the learning of high-frequency information within radiance fields. Specifically, given an input coordinate \({t} \in \mathbb{R}\), NeRF maps it into \(\mathbb{R}^{2L}\) space as follows,
\begin{equation}
\label{eq:pe_neff}
\begin{aligned}
    &\gamma_{\text{NeFF}}({t}) = \\ &\left[\sin (2^0 \pi {t}), \cos (2^0 \pi {t}), \ldots, \sin (2^{L-1} \pi {t}), \cos (2^{L-1} \pi {t})\right]
\end{aligned}
\end{equation} where \(L\) represents the dimension of the high-frequency space. We refer to this positional encoding scheme as \texttt{NeFF} (NeRF Fourier Feature).

\begin{table*}[htbp]
    \centering
\fontsize{9}{11}\selectfont
  \begin{tabular}{c|c|ccccccc}
    \cline{1-9}

    & \multicolumn{2}{c}{\textbf{Parameter} $L$} & 2 &  4 & 8  & 16 & 32 & 64 \\
    \cdashline{2-9}
     Dataset & \multicolumn{2}{c}{Params} & 264K & 265K  &  268K & 272K &  280K & 296K  \\
    
    \cline{1-9}
   \multirow{8}{*}{Bach(\texttt{7s})} & \multirow{2}{*}{Gaussian} & SNR $\uparrow$ & 16.12 & \fs 29.08  & \underline{ 25.16} & 19.86 & 21.06 & \textit{ 23.48}  \\
    & & LSD $\downarrow$ & \underline{ 0.810} & \fs 0.737 & \textit{ 1.785} & 2.115 & 2.035 &1.880 \\ 

  \cdashline{2-9}
   &\multirow{2}{*}{ReLU} & SNR $\uparrow$ &0.00 & 0.02 & 4.60 & \fs 24.76 & \underline{ 5.37} & \textit{ 4.89} \\
    & & LSD $\downarrow$ & 3.997 & 3.590 & \fs 1.151 & \underline{ 1.590} & \textit{ 2.812 }& 3.000  \\ 

    \cdashline{2-9}
   &\multirow{2}{*}{Sine} & SNR $\uparrow$ & 12.12 & 14.75 & \textit{ 34.63} & \fs 46.40 & 34.36 & \underline{ 37.62} \\
    & & LSD $\downarrow$ & 0.914 & \textit{ 0.864} & \underline{ 0.663} & \fs 0.530 &1.328 & 1.188  \\ 

    \cdashline{2-9}
   &\multirow{2}{*}{Incode-Sine} & SNR $\uparrow$ & 14.25 & 15.70 & 33.38 & \underline{ 36.94} & \textit{ 33.46} & \fs 37.53  \\
    & & LSD $\downarrow$ & \underline{ 0.814 }& \textit{ 0.837} & \fs 0.663 & 1.120&1.332 & 1.145  \\ 
  \hline
  \hline

   \cline{1-9}
   \multirow{8}{*}{Counting(\texttt{7s})} & \multirow{2}{*}{Gaussian} & SNR $\uparrow$ &7.50  & 9.89 & \fs 19.31& 9.36 &\underline{  10.71} & \textit{ 10.66}  \\
    & & LSD $\downarrow$ & \underline{ 1.760} & \fs 1.567 & \textit{ 2.589} & 3.429 & 3.323 & 3.221 \\

  \cdashline{2-9}
   &\multirow{2}{*}{ReLU} & SNR $\uparrow$ & 0.00 & 0.01 & 0.59 &\fs  13.27 & \underline{ 6.64} &  \textit{ 6.47} \\
    & & LSD $\downarrow$ & 3.548 & 3.256 & \underline{ 1.938} &\fs 1.326 & \textit{ 2.618} & 2.812   \\

    \cdashline{2-9}
   &\multirow{2}{*}{Sine} & SNR $\uparrow$ & 7.04 & 10.36 & 12.73 & \fs 35.46 & \textit{ 24.16} & \underline{ 27.08} \\
    & & LSD $\downarrow$ & \textit{ 1.723 }& 1.803 & \underline{ 1.713} & \fs 1.377 & 2.216 & 1.979 \\

    \cdashline{2-9}
   &\multirow{2}{*}{Incode-Sine} & SNR $\uparrow$ & 8.65 & 11.03  &12.55 & \fs 28.40 & \textit{ 13.06} & \underline{ 14.14}  \\
    & & LSD $\downarrow$ & \fs 1.649 & \textit{ 1.784} & \underline{ 1.752} &1.855 & 3.145& 3.056  \\ 

    \hline
    \hline

   \cline{1-9}
   \multirow{8}{*}{Blues(\texttt{30s})} & \multirow{2}{*}{Gaussian} & SNR $\uparrow$ & 3.12 & \underline{ 8.84} & \fs 19.82 & 7.81 & 8.50 & \textit{ 8.78}  \\
    & & LSD $\downarrow$ &2.057 & 1.580 & \fs 0.849 & 1.594 & \textit{ 1.553} & \underline{ 1.536} \\

  \cdashline{2-9}
   &\multirow{2}{*}{ReLU} & SNR $\uparrow$ & 0.00  & 0.01 & 0.67& \fs 14.74& \textit{ 2.58} & \underline{ 2.81}  \\
    & & LSD $\downarrow$ & 5.140 & 4.708 & 2.843 &\fs 0.890 & \underline{ 1.730} & \textit{ 1.760 }\\

    \cdashline{2-9}
   &\multirow{2}{*}{Sine} & SNR $\uparrow$ & 7.39 & 9.76 & \underline{ 12.96} & \fs 24.08 & 10.80& \textit{ 11.68} \\
    & & LSD $\downarrow$ & 1.697 & 1.578  & \underline{ 1.360} & \fs 0.606 & 1.422 &  \textit{ 1.370} \\

    \cdashline{2-9}
   &\multirow{2}{*}{Incode-Sine} & SNR $\uparrow$ &6.81  & 8.22 & \underline{ 11.90} & \fs 22.86 &10.26 & \textit{ 11.09} \\
    & & LSD $\downarrow$ & 1.751 & 1.667 & \textit{ 1.449} & \fs 0.670 & 1.454 & \underline{ 1.405} \\ 

    \hline
  \end{tabular}
 \caption{The impact of the parameter $L$ in \texttt{NeFF} positional encoding (Eq. \ref{eq:pe_neff}) on the quality of audio representation. For each row, best results are highlighted as  \textbf{first}, \underline{second},  and \textit{third}. }
  \label{tab:neff_l}
\end{table*}

\begin{table*}[!htbp]
    \centering
\fontsize{9}{11}\selectfont
  \begin{tabular}{c|c|ccccccccc}
    \cline{1-11}

    & \multicolumn{2}{c}{\textbf{Parameter} $L$} & 2 &  4 & 8  & 16 & 32 & 64 & 128 & 256 \\
    \cdashline{2-11}
     Dataset & \multicolumn{2}{c}{Params} & 264K & 265K  & 267K  & 271K & 280K & 296K & 329K & 394K \\
    
    \cline{1-11}
   \multirow{8}{*}{Bach(\texttt{7s})} &\multirow{2}{*}{Gaussian} & SNR $\uparrow$ & 18.61 & 19.41 & 19.67 & 20.21 & 20.84 & \textit{ 23.89} & \underline{ 27.06} & \fs 30.18 \\
    & & LSD $\downarrow$ &2.199 & 2.143 & 2.125 & 2.090 & 2.046 & \textit{ 1.850} & \underline{ 1.680} & \fs 1.527 \\ 
    
    \cdashline{2-11}
    & \multirow{2}{*}{ReLU} & SNR $\uparrow$ & 0.82 & 13.44 & 14.57 & 14.87 &  \fs 15.62 & 14.55 & \textit{ 14.94} & \underline{ 15.08} \\
    & & LSD $\downarrow$ & 1.580 & 1.088 & 1.019 & 0.988 & 0.978 & \textit{ 0.951} &  \underline{ 0.932} & \fs 0.909 \\ 

    \cdashline{2-11}
    & \multirow{2}{*}{Sine} & SNR $\uparrow$ & \fs 43.24 & \underline{ 41.97} & \textit{ 41.60} & 40.04 &39.02  & 38.27 & 38.37 &  38.82 \\
    & & LSD $\downarrow$ & 0.778 & \underline{ 0.548} & \fs 0.547 & \textit{ 0.561} & 0.582 & 0.595 & 0.599 & 0.580 \\ 

     \cdashline{2-11}
    & \multirow{2}{*}{Incode-Sine} & SNR $\uparrow$ & \fs 41.82 & \underline{ 40.01} & \textit{ 39.84} & 39.02 & 38.10  & 37.17 & 36.71 & 36.94 \\
    & & LSD $\downarrow$ & 0.877 & \fs 0.567 & \underline{ 0.568} & \textit{ 0.575} & 0.595 & 0.614 & 0.627 & 0.610 \\ 

      \hline
      \hline

   \multirow{8}{*}{Counting(\texttt{7s})} &\multirow{2}{*}{Gaussian} & SNR $\uparrow$ & 8.73 & 9.37 & 9.69 & 10.33 & 12.14 & \fs 20.09 & \underline{ 16.78} & \textit{ 16.34}\\
    & & LSD $\downarrow$ & 3.474 & 3.424 & 3.400 & 3.347 & 3.195 & \underline{ 2.335} & \fs 2.119  & \textit{ 2.445}\\
    
    \cdashline{2-11}
    & \multirow{2}{*}{ReLU} & SNR $\uparrow$ & 0.74 & 4.09 & \underline{ 4.52} & \textit{ 4.48} & \fs 4.93 &  4.22 & 4.28 & 4.42\\
    & & LSD $\downarrow$ & 1.690 & 1.489 & 1.430 & 1.412 & 1.400 &\textit{ 1.379}  & \underline{ 1.361} & \fs 1.354  \\ 

     \cdashline{2-11}
     & \multirow{2}{*}{Sine} & SNR $\uparrow$ & \fs 28.64 & \underline{ 25.76} & \textit{ 17.62} &13.50 & 13.05  &12.97  & 13.02 & 12.92 \\
    & & LSD $\downarrow$ & 1.396 & \fs 1.077 & \underline{ 1.184} & \textit{ 1.253} & 1.412 & 1.439 & 1.434 & 1.571 \\ 

     \cdashline{2-11}
    & \multirow{2}{*}{Incode-Sine} & SNR $\uparrow$ & \fs 27.59 & \underline{ 19.26} & \textit{ 13.34} &12.95 &12.86  & 12.83 & 12.81 & 12.80\\
    & & LSD $\downarrow$ &\textit{ 1.332} & \fs 1.203 & \underline{ 1.269} & 1.472 & 1.558 & 1.649 & 1.648 & 1.682 \\ 

    \hline
    \hline
    \multirow{8}{*}{Blues(\texttt{30s})} &\multirow{2}{*}{Gaussian} & SNR $\uparrow$ & 7.74 & 8.57  & 8.89 &9.66 &  11.74& \textit{ 18.06} & \underline{ 21.98} & \fs 23.16 \\
    & & LSD $\downarrow$ & 1.593 & 1.541 & 1.522 & 1.475 & 1.348& \textit{ 0.975} & \underline{ 0.763}  & \fs 0.672 \\
    
    \cdashline{2-11}
    & \multirow{2}{*}{ReLU} & SNR $\uparrow$ &  0.56 & 2.57 &2.88  & 2.91 & \fs 3.23 & 2.83 & \underline{ 2.92} & \textit{ 2.90}\\
    & & LSD $\downarrow$ & 2.343& 1.964 & \underline{ 1.913} & 1.920 & \fs 1.855 & 1.950 & 1.920 & \textit{ 1.919} \\ 

     \cdashline{2-11}
    & \multirow{2}{*}{Sine} & SNR $\uparrow$ & \fs 21.13 & \textit{ 19.91} & 19.01 & 17.68 & 17.08 & 17.29 & 18.35 & \underline{ 20.49} \\
    & & LSD $\downarrow$ &\fs 0.742 & \underline{ 0.908} & \textit{ 0.986} & 1.078 & 1.130 & 1.103 & 1.035 & 0.874 \\ 

     \cdashline{2-11}
    & \multirow{2}{*}{Incode-Sine} & SNR $\uparrow$ & \fs 21.09 & \underline{ 19.48} & \textit{ 18.37} & 16.77 & 15.60 & 15.11 & 14.14 & 16.71 \\
    & & LSD $\downarrow$ & \fs 0.765 & \underline{ 0.953} & \textit{ 1.047} & 1.445 & 1.215 & 1.123 & 1.172 & 1.137  \\ 
  \hline
  \end{tabular}
 \caption{The impact of the parameter $L$ in \texttt{RFF} positional encoding (Eq. \ref{eq:pe_rff}) on the quality of audio representation.}
  \label{tab:rff_l}
\end{table*}

According to the neural tangent kernel (NTK) theory, FFN \cite{tancik2020ffn} introduces Gaussian random noise into the NeRF positional encoding (denoted as \texttt{RFF}, Random Fourier Feature), as follows,
\begin{equation}
\label{eq:pe_rff}
\begin{aligned}
&\gamma_{\text{RFF}}({t})=\\ &\left[\cos (2 \pi \mathbf{b}_1^{\mathrm{T}} {t}), \sin (2 \pi \mathbf{b}_1^{\mathrm{T}} {t}), \ldots, \cos (2 \pi \mathbf{b}_L^{\mathrm{T}} {t}), \sin (2 \pi \mathbf{b}_L^{\mathrm{T}} {t})\right]^{\mathrm{T}}, \\
& \qquad \qquad \qquad \mathbf{b}_i \sim \mathcal{N}(0, \sigma^2),
\end{aligned}
\end{equation} where frequency vectors $\mathbf{b}_i$ are randomly sampled from a Gaussian distribution with a standard deviation of $\sigma$ (hyper-parameter). This mapping transforms the NTK into a stationary (shift-invariant) kernel and enables tuning the NTK's spectrum by modifying the frequency vectors $\mathbf{b}_i$, thereby controlling the range of frequencies that can be learned by the corresponding MLP.

These encoding schemes facilitate the learning of complex functions but also introduce a secondary phenomenon: a reduction in the network's spectral bias \cite{ramasinghe2022regularizingcoordinatemlps}. The implications of this situation are twofold. Positional encoding can lead to instability in the neural optimization process, potentially causing convergence to suboptimal local minima. Furthermore, when fitting functions with varying levels of detail or frequency, neurons predicting smooth, low-frequency regions are still exposed to high-frequency encoding dimensions. This exposure can complicate the learning process, as the network must learn when and how to disregard these embedding dimensions. Indeed, FFN \cite{tancik2020ffn} argues for careful selection of the frequency distribution's standard deviation $\sigma$ in \texttt{RFF} positional encoding, noting that low values may result in a lack of detail in the network output while excessively high values can introduce noisy artifacts. Therefore, the value of $\sigma$ for such static encoding needs to be fine-tuned for each sample.

\textbf{Experiments.}
Table \ref{tab:neff_l} presents an experiment on the sensitivity of the dimensionality parameter \( L \) for the \texttt{NeFF} (Eq. \ref{eq:pe_neff}) positional encoding. It is evident that the optimal value of \( L \) varies depending on the activation function used. Moreover, it is noteworthy that there is a significant difference in the audio reconstruction results between the optimal and suboptimal parameters. For instance, in the case of the ReLU activation function, the optimal parameter \( L=16 \) achieves an SNR of 24.76 dB, whereas the suboptimal parameter \( L=32 \) results in an SNR of only 5.37 dB. This substantial disparity complicates the parameter tuning process.

Tables \ref{tab:rff_l} and \ref{tab:rff_sigma} present experiments on the sensitivity of the \texttt{RFF} positional encoding to the dimensionality parameter \( L \) and the variance parameter \( \sigma \), respectively. The results indicate that \texttt{RFF} is more sensitive to the dimensionality parameter \( L \), as observed from the distribution of the optimal, second-best, and third-best configurations. For the Gaussian activation function, the optimal SNR is achieved at \( L=256 \), although this also results in a higher parameter count. For the ReLU activation function, the optimal value is \( L=32 \). In contrast, for periodic activation functions (e.g., Sine, Incode-Sine), the optimal value is \( L=2 \), yet there is a significant disparity in reconstruction quality between the optimal and suboptimal parameters.

According to Table \ref{tab:rff_sigma}, although the variance parameter \( \sigma \) that yields the highest SNR tends to be around 1000, it is important to note that a higher \( \sigma \) does not necessarily lead to better results. In fact, the suboptimal values of \( \sigma \) are often less than 1000. This variation complicates the parameter tuning process, making it challenging to identify the optimal settings.

\begin{table*}[!ht]
    \centering
\fontsize{9}{11}\selectfont
  \begin{tabular}{c|c|ccccccccc}
    \cline{1-11}

    Dataset & \multicolumn{2}{c}{\textbf{Parameter} $\sqrt{\sigma}$} & 1 &  20 & 40 & 60 & 80 & 100 & 1000 & 10000  \\
    
    \cline{1-11}
   \multirow{8}{*}{Bach(\texttt{7s})} &\multirow{2}{*}{Gaussian} & SNR $\uparrow$ & 19.02 & \fs 27.72 & \underline{ 23.21} & \textit{ 21.68}  & 21.12 & 20.84 & 20.58 & 20.63  \\
    & & LSD $\downarrow$ & \fs 0.801 & \underline{ 1.628} & \textit{ 1.899} & 1.993 & 2.028 & 2.046 & 2.064 & 2.063 \\ 
    
    \cdashline{2-11}
    & \multirow{2}{*}{ReLU} & SNR $\uparrow$ & 0.01 & 4.42 & 9.69 & 11.91 & \textit{ 13.95} & \underline{ 15.62} & \fs 30.30 & 5.99 \\
    & & LSD $\downarrow$ &3.968  & 1.165 & 0.988 & \underline{ 0.977} & \fs 0.970 & \textit{ 0.978} &  1.374 & 3.011 \\ 

    \cdashline{2-11}
    & \multirow{2}{*}{Sine} & SNR $\uparrow$ & 12.00  & 28.38 & 34.43 & 36.92 & \textit{ 38.16}  & \underline{ 39.02} & \fs 45.27 & 33.23 \\
    & & LSD $\downarrow$ &0.920 & 0.661 & 0.636 & 0.612 & \textit{ 0.592} & \underline{ 0.582} & \fs 0.517 & 1.344 \\

   \cdashline{2-11}
    & \multirow{2}{*}{Incode-Sine} & SNR $\uparrow$ & 13.71 & 24.91 & 33.03 & 36.04 & \textit{ 37.30}  & \underline{ 38.10 }& \fs 46.35 & 33.49 \\
    & & LSD $\downarrow$ &0.859 & 0.694 & 0.631 & 0.624  & \textit{ 0.607 }& \underline{ 0.595} & \fs 0.516  & 1.330 \\ 
   
      \hline
      \hline
      
   \multirow{8}{*}{Counting(\texttt{7s})} &\multirow{2}{*}{Gaussian}  & SNR $\uparrow$ & 8.12 & \fs  20.95 & \underline{ 16.45} & \textit{ 14.11} &  13.24 & 12.14 & 10.31 & 10.37 \\
    & & LSD $\downarrow$ & \fs 1.651 & \underline{ 2.327} & \textit{ 2.849} & 3.041 & 3.108 & 3.195 & 3.353 & 3.349 \\ 
    
    \cdashline{2-11}
    & \multirow{2}{*}{ReLU} & SNR $\uparrow$ & 0.00 & 0.589 & 1.40 & 2.61 &3.82 & \textit{ 4.93} & \fs 13.21 & \underline{ 8.58} \\
    & & LSD $\downarrow$ & 3.417 & 1.964 & 1.630 & \textit{ 1.466} & \underline{ 1.402} & \fs 1.400 & 1.939  & 3.211  \\ 

    \cdashline{2-11}
    & \multirow{2}{*}{Sine} & SNR $\uparrow$ & 6.29 & 12.63 & 12.75 & 12.85  & 12.91  &  \textit{ 13.06} & \fs 35.13 & \underline{ 22.59} \\
    & & LSD $\downarrow$ & 1.666 & 1.706 & 1.685 & 1.640 & \textit{ 1.532}  & \underline{ 1.412} & \fs 1.113 & 2.364 \\

   \cdashline{2-11}
    & \multirow{2}{*}{Incode-Sine} & SNR $\uparrow$ & 6.23 & 12.13 & 12.56 & 12.78 & 12.84  &\textit{  12.86} & \fs 34.65  & 
    \underline{ 16.09} \\
    & & LSD $\downarrow$ & 1.598 & 1.762 & 1.702 & 1.652 & \textit{ 1.585} & \underline{ 1.559} & \fs 1.114  & 2.890 \\ 

    \hline
    \hline
   \multirow{8}{*}{Blues(\texttt{30s})} &\multirow{2}{*}{Gaussian}  & SNR $\uparrow$ & 4.85 & \fs 21.37 & \underline{ 17.33} & \textit{ 14.19} & 12.59  & 11.74 &8.56 & 8.57 \\
    & & LSD $\downarrow$ & 1.876 &\fs  0.920 &\underline{ 1.026} & 1.211 & \textit{ 1.302 }& 1.348 & 1.549 & 1.549  \\ 
    
    \cdashline{2-11}
    & \multirow{2}{*}{ReLU} & SNR $\uparrow$ & 0.00  & 0.66 & 1.39 & 2.08 & 2.68   &\textit{ 3.23}  & \fs 13.58 &  \underline{ 9.05}  \\
    & & LSD $\downarrow$ & 4.893 & 2.826 & 2.386 & 2.136 & 1.986  &\textit{ 1.855}  & \fs 0.949 & \underline{ 1.374} \\ 

    \cdashline{2-11}
    & \multirow{2}{*}{Sine} & SNR $\uparrow$ &4.91  &15.21  & 14.29 &15.40  & 16.22  & \underline{ 17.09} & \fs 23.53 & \textit{ 16.78}  \\
    & & LSD $\downarrow$ &1.895 & 1.235 & 1.267 & 1.215 & 1.171 & \textit{ 1.130} & \fs 0.583 & \underline{ 1.084} \\

   \cdashline{2-11}
    & \multirow{2}{*}{Incode-Sine} & SNR $\uparrow$ & 3.866 & 10.75 & 12.08 &13.14  &  \textit{ 14.32} & \underline{ 15.60} & \fs 22.86 & 13.88  \\
    & & LSD $\downarrow$ & 1.977 & 1.542 & 1.426 & 1.366 & 1.277 & \underline{ 1.215} & \fs 0.619 & \textit{ 1.250}  \\ 
  \hline
  \end{tabular}
\caption{The impact of the parameter $\sigma$ in \texttt{RFF} positional encoding (Eq. \ref{eq:pe_rff}) on the quality of audio representation. For each row, best results are highlighted as  \textbf{first}, \underline{second},  and \textit{third}.}
  \label{tab:rff_sigma}
\end{table*}

\subsection{A.2 Activation functions are parameter-sensitive}
\textbf{Analysis.} Our benchmark leaderboard indicates that most activation functions are unable to capture the high-frequency and local periodic characteristics of audio signals. However, effective activation functions incorporate a scaling factor hyper-parameter to represent signals at different frequencies. As shown in Table \ref{tab:hyper_activations}, Gaussian-type activation functions utilize a variance factor \(a\), while sine-type activation functions use a frequency factor \(\omega\). Both types require careful parameter tuning to effectively represent audio signals.

\begin{table}[!h]
    \centering
    \fontsize{9}{11}\selectfont
    \begin{tabular}{c|c|c}
        \hline 
        Activation Function & Equation & Parameter \\
        \hline 
        Gaussian & $e^{\frac{-x^2}{2 a^2}}$ & $[a]$ \\
        \hline
        Laplacian & $e^{\frac{-|x|}{a}}$ & $[a]$ \\
        \hline
        
        Super-Gaussian & $[e^{\frac{-x^2}{2 a^2}}]^b$ & $[a, b]$ \\
        
        \hline 
        Sine & $\sin (\omega x)$ & $[\omega]$ \\
        
        \hline 
        Incode-Sine & $a \sin (b \omega x+c)+d$ & $[\omega], a, b, c,d$ \\
        \hline
    \end{tabular}
     \caption{Activation functions with hyper-parameters.}
    \label{tab:hyper_activations}
\end{table}

\textbf{Experiments.} 
Table \ref{tab:a_gaussian} presents experiments on the sensitivity of the Gaussian activation function to the variance parameter \( a \). The results indicate that the variance of the Gaussian function significantly influences the representation of audio signals, leading to distinct optimal values of the parameter \( a \) for different audio signals. This variability necessitates intricate parameter tuning, making the adjustment process more complex.
\begin{table}[!ht]
    \centering
  \tabcolsep=1.0mm
 \fontsize{9}{11}\selectfont
  \begin{tabular}{c|ccccccc}
    \hline
     Dataset & Parameter $a$ & 0.01 &  0.1 & 0.2 & 0.3 & 1.0   \\
    
   \hline
   \multirow{2}{*}{Bach(\texttt{7s})} & SNR $\uparrow$ & 0.01 & \fs 6.35 & 0.04 & 0.00 & -0.01  \\
    & LSD $\downarrow$ & 1.799 & \fs 1.130 & 3.851 & 4.120 & 6.266 \\ 

    \hline
    \multirow{2}{*}{Counting(\texttt{7s})} & SNR $\uparrow$ & \fs 0.74 & 0.34 & 0.01 & 0.00 & -0.01  \\
    & LSD $\downarrow$ & \fs 1.889 & 2.165 & 3.334 & 4.107 & 5.169  \\ 
  \hline
  \end{tabular}
\caption{The impact of the variance parameter \( a \) on Gaussian activation function. For each row, best results are highlighted as  \textbf{first}.}
  \label{tab:a_gaussian}
\end{table}

Table \ref{tab:omega_incode_sine} presents experiments assessing the sensitivity of the frequency hyperparameter \( \omega \) for the periodic activation functions Sine and Incode-Sine under various positional encoding settings. The results clearly demonstrate that positional encoding mitigates the sensitivity of periodic activation functions to frequency, with the optimal frequency parameter converging around 30. This can be attributed, in part, to the frequency-dependent initialization strategy employed by periodic activation functions (see A.3), leading the optimal frequency to align closely with the frequency used in the initialization strategy. Additionally, Incode-Sine reduces the sensitivity of the Sine activation function to frequency by making the parameters within the Sine function learnable.

\begin{table*}[!ht]
    \centering
    \fontsize{9}{11}\selectfont
  \begin{tabular}{c|c|ccccccc}
    \hline
    Dataset & \multicolumn{2}{c}{\textbf{Parameter} $\omega$} & 3 &  30 & 300 & 3,000 & 30,000 & 300,000   \\
    
    \cline{1-9}
   \multirow{10}{*}{Bach(\texttt{7s})} &\multirow{2}{*}{Incode-Sine} & SNR $\uparrow$ & -1.0 &  \fs 15.98 & 0.00 & -0.01 & -0.02 & -0.01  \\
    & & LSD $\downarrow$ &6.341 & \fs 0.778 & 2.441 & 2.442 & 2.432 & 2.466  \\ 

    \cdashline{2-9}
    & \multirow{2}{*}{\texttt{RFF}+Incode-Sine} & SNR $\uparrow$ & 10.97 & \fs 38.10 & 0.05 & -0.02 & -0.01 & 0.01  \\
    & & LSD $\downarrow$ & 0.997 & \fs 0.595 & 2.465 & 2.389 & 2.491 & 2.483 \\ 
    
    \cdashline{2-9}
    & \multirow{2}{*}{\texttt{NeFF}+Incode-Sine} & SNR $\uparrow$ & 0.00 & \fs 15.70 & 0.00 & -0.01 &  0.00 & -0.02  \\
    & & LSD $\downarrow$ & 5.685 & \fs 0.837 & 2.442 & 2.446 & 2.464 & 2.419    \\ 
    
    \cdashline{2-9}
    & \multirow{2}{*}{Sine} & SNR $\uparrow$ & 0.30 &  12.27 & 23.96 & 39.59 & \fs 40.36 & 0.01 \\
    & & LSD $\downarrow$ & 2.478 & 0.885 & 0.669 & \fs 0.575 &  0.683 & 2.526  \\ 

     \cdashline{2-9}
    & \multirow{2}{*}{\texttt{NeFF}+Sine} & SNR $\uparrow$ & 4.13 & \fs 42.39 & 0.80 & 0.70 & 1.12 & 0.00 \\
    & & LSD $\downarrow$ & 1.348 & \fs 0.537 & 2.537 & 2.589 & 2.523 & 2.678  \\ 

    \hline
    \hline
    
   \multirow{10}{*}{Counting(\texttt{7s})} &\multirow{2}{*}{Incode-Sine} & SNR $\uparrow$ & -0.05 &  \fs 8.16 & -0.02 & -0.02 &  -0.06 & -0.01\\
    & & LSD $\downarrow$ & 5.184 & \fs 1.611 & 2.116 & 2.100 & 2.126 & 2.139 \\ 

    \cdashline{2-9}
    & \multirow{2}{*}{\texttt{RFF}+Incode-Sine} & SNR $\uparrow$ & 4.04 & \fs 12.86 & -0.09 & -0.07 & -0.08 & -0.03 \\
    & & LSD $\downarrow$ & 1.730 & \fs 1.559 & 2.138 & 2.114 & 2.106 & 2.125\\ 
    
    \cdashline{2-9}
    & \multirow{2}{*}{\texttt{NeFF}+Incode-Sine} & SNR $\uparrow$ &  -0.03 & \fs 11.03 &  -0.09 & -0.07 & -0.06 &  -0.01 \\
    & & LSD $\downarrow$ & 5.167 & \fs 1.784 & 2.104 & 2.126 & 2.121 & 2.114 \\ 

    \cdashline{2-9}
    & \multirow{2}{*}{Sine} & SNR $\uparrow$ & 0.26 & 7.98 & 11.79 & \fs 12.94 & 0.40 & -0.01  \\
    & & LSD $\downarrow$ & 2.153 & 1.716 & 1.710 & \fs 1.533 & 2.120 & 2.129 \\ 

     \cdashline{2-9}
    & \multirow{2}{*}{\texttt{NeFF}+Sine} & SNR $\uparrow$ & 4.81 & \fs 33.57 &  0.82 & 0.89 & 0.27 & 0.06 \\
    & & LSD $\downarrow$ & 1.849 & \fs 0.914 & 2.135 & 2.124 & 2.145 & 2.123 \\ 

  \hline
  \end{tabular}
\caption{The impact of the frequency parameter \( \omega \) on Sine and Incode-Sine activation functions. For each row, best results are highlighted as  \textbf{first}.}
  \label{tab:omega_incode_sine}
\end{table*}

\subsection{A.3 Periodic activation functions are sensitive to initialization schemes}
\textbf{Analysis.} Since 1987, periodic activation functions have been extensively studied, beginning with the work of Lapedes and Farber \cite{lapedes1987periodic}, which demonstrates that networks with such activations are generally difficult to train. Parascandolo et al. \cite{parascandolo2017taming} further investigated the challenges of training networks with periodic activation functions. They noted that training is successful only when the network does not rely on the periodicity of the given function, and they recommended using truncated sine functions. Inspired by the discrete cosine transform, Klocek et al. \cite{Klocek2019Hypernetwork} proposed using cosine activation functions to target a network whose weights are determined by a hypernetwork.

Recently, SIREN \cite{sitzmann2020siren} demonstrates that MLPs with sine activation functions are naturally well-suited for encoding high-frequency signals, thereby eliminating the need for positional encoding layers. Despite their potential, researches involving Coordinate-MLPs tend to focus more on positional encoding than on sine activations. This trend may be attributed to two primary reasons. First, Sitzmann et al. mainly attribute the success of sine activations to their periodicity, although there is insufficient evidence to support this relationship. This lack of understanding obscures some fundamental principles of their effectiveness, hindering their reliable application in broader contexts \cite{ramasinghe2022activations}. Second, the performance of MLPs with sine activations is highly sensitive to initialization. When MLP initialization does not strictly adhere to the guidelines set forth by Sitzmann et al., performance can significantly deteriorate.

Specifically, models such as SIREN \cite{sitzmann2020siren} and INCODE \cite{kazerouni2023incode}, which utilize sine activation functions ($\sigma(x)=\sin(\omega x)$), necessitate the adoption of \textbf{Sitzmann's initialization} scheme. For a linear combination of $n$ inputs $\mathbf{x} \in \mathbb{R}^n$ with weights $\mathbf{w} \in \mathbb{R}^n$, the output is given by $y=\sin \left(\omega \mathbf{w}^T \mathbf{x}+\mathbf{b}\right)$. If the neuron is in the input layer, each component of $\mathbf{w}$ must follow a uniform distribution over $[-1,1]$, 
$$
w_i \sim \mathcal{U}(-\frac{1}{\sqrt{n}}, \frac{1}{\sqrt{n}}).
$$ For neurons not in the input layer, each component of $\mathbf{w}$ must follow a distribution related to the frequency $\omega$, 
$$
w_i \sim \mathcal{U}(-\frac{c}{\omega \sqrt{n}}, \frac{c}{\omega \sqrt{n}}),
$$ where \(c\) is an additional parameter, which SIREN sets to 6. 

Due to the dependency of this initialization scheme on the frequency hyperparameter $\omega$, and the lack of generalizability of the frequency parameter, sine-type activation functions are highly sensitive to the initialization scheme.
\begin{table}[!htbp]
    \centering
  \tabcolsep=1.0mm
   \fontsize{9}{11}\selectfont

  \begin{tabular}{c|cc|cc|cc}
    \cline{1-7}
     \multirow{2}{*}{Initialization} &  \multicolumn{2}{c}{Bach} &  \multicolumn{2}{c}{Counting} & \multicolumn{2}{c}{Blues} \\
    \cline{2-7}
      & SNR $\uparrow$ & LSD $\downarrow$  & SNR $\uparrow$ & LSD $\downarrow$ & SNR $\uparrow$ & LSD $\downarrow$ \\
    \cline{1-7}
    Kaiming &2.5-e5 & 2.471 & -6.9-e5 & 2.144 & -3.4-e5 & 3.451 \\

    \cdashline{1-7}
    Xvavier & -3.7-e5 & 4.42 & 3.3-e5 & 2.171 & 1.3-e5  & 3.167 \\

    \cdashline{1-7}
    Sitzmann & \fs 13.36 & \fs 0.838 & \fs 7.96 & \fs 1.660 & \fs  7.47 & \fs 1.722 \\
    
  \cline{1-7}
  \end{tabular}
 \caption{Periodic activation functions are sensitive to initialization schemes.}
  \label{tab:periodic_sensitive}
\end{table}

\textbf{Experiments.} As shown in Table \ref{tab:periodic_sensitive}, we evaluate the performance of Sine (Table \ref{tab:hyper_activations}) activation function for representing audio signals using Kaiming initialization \cite{he2015kaiminginit}, Xavier initialization \cite{pmlr2010xavierinit}, and Sitzmann initialization \cite{sitzmann2020siren}. The results clearly indicate that without the frequency-dependent initialization proposed by Sitzmann, Sine-type periodic activation functions fail to converge and are unable to represent audio signals.

\section{B. Memorization}
Since implicit neural representations encode discrete signals as neural network weights, the effectiveness of the network is largely contingent upon its ability to memorize the training data. In other words, this effectiveness is determined by the network's capacity to represent the complexity of the signals. Consequently, this section primarily analyzes the representational capacity of Coordinate-MLPs and Kolmogorov–Arnold networks, including the Fourier-KAN proposed in our study.

\subsection{B.1 Memorization of Coordinate-MLPs}
\label{memorization_mlps}
Beyond Periodicity \cite{ramasinghe2022activations} demonstrates that the effectiveness of Coordinate-MLPs is closely related to their Lipschitz smoothness and the eigenvalue distribution of the hidden layer representations. The optimal values of these metrics are contingent upon the characteristics of the signals to be encoded. We will outline the main proof process below.

\textbf{Derivation.} A multilayer perceptron (MLP) \( f \) with \( k-1 \) nonlinear hidden layers can be described as follows,
$$
f: \mathbf{x} \rightarrow g^k \circ \psi^{k-1} \circ g^{k-1} \circ \cdots \circ \psi^1 \circ g^1(\mathbf{x}),
$$ where \( g^i: \mathbf{x} \rightarrow \mathbf{A}^i \cdot \mathbf{x} + \mathbf{b}^i \) represents an affine transformation with trainable weights, \(\mathbf{A}^i \in \mathbb{R}^{\operatorname{dim}(\mathbf{x}^i) \times \operatorname{dim}(\mathbf{x}^{i-1})}\) is the weight matrix, \(\mathbf{b}^i \in \mathbb{R}^{\operatorname{dim}(\mathbf{x}^i)}\) is the bias term, and \(\psi^i\) is a nonlinear activation function. The final layer is a linear transformation, yielding \( f: \mathbf{x} \rightarrow g^k \circ \phi(\mathbf{x}) \), where \(\phi\) is the composite function formed by the first \( k-1 \) layers of the MLP, excluding the final linear layer. Given \( N \) training samples, the total (training) embedding matrix can be defined as,
$$
\mathbf{X} \in \mathbb{R}^{D \times N} := \left[\phi(\mathbf{x}_1)^T \; \phi(\mathbf{x}_2)^T \; \ldots \; \phi(\mathbf{x}_N)^T\right],
$$ where \(\{\mathbf{x}_n\}_{n=1}^N\) represents the original training inputs.

It is important to note that the final layer of an MLP is typically an affine projection without any nonlinearity. For simplicity in notation, the bias term can be omitted, resulting in,
$$
\tilde{\mathbf{Y}} = \mathbf{A}^k \mathbf{X},
$$ where \(\tilde{\mathbf{Y}} \in \mathbb{R}^{q \times N}\) denotes the output of the MLP.

Assume that \(\mathbf{Y} \in \mathbb{R}^{q \times N}\) represents the true training outputs that the MLP aims to learn. If the MLP perfectly memorizes the training set, i.e., \(\tilde{\mathbf{Y}} = \mathbf{Y}\), then each row of \(\mathbf{Y}\) can be expressed as a linear combination of the rows of \(\mathbf{X}\). Assuming no prior knowledge of \(\mathbf{Y}\), the rows of \(\mathbf{Y}\) can be arbitrary vectors in \(\mathbb{R}^N\). If the rows of \(\mathbf{X}\) are linearly independent, they form a basis for \(\mathbb{R}^N\) (assuming \(D \geq N\)). Therefore, if \(\operatorname{rank}(\mathbf{X}) = N\), it is guaranteed (assuming perfect convergence) that the MLP can identify a weight matrix \(\mathbf{A}^k\) to ensure the perfect reconstruction of \(\mathbf{Y}\).

It is important to note that while the condition \(\operatorname{rank}(\mathbf{X}) = N\) is sufficient to ensure perfect memorization of any signal, it may not always be necessary. This is because natural signals often exhibit redundancy—i.e., they have limited bandwidth. The bandwidth of a signal class can be defined as the number of linearly independent (normalized) basis vectors required to represent them. Consequently, for many signal classes, \(\operatorname{rank}(\mathbf{X})\) can be less than \(N\) while the MLP can still achieve perfect signal reconstruction. Furthermore, the stable rank provides a lower bound for the rank \cite{Rudelson2007rank}. A higher stable rank generally leads to better reconstruction performance, though this metric is constrained by the network width (\(D\)), which is less than the number of data points (\(N\)). Conversely, encoding noisy signals with limited or no redundancy necessitates a larger network width, and poorer results are typically observed when \(D \ll N\).

\textbf{Analysis.} Based on the aforementioned derivation, we can briefly analyze the audio representation capabilities of Gaussian-MLPs and ReLU-MLPs in terms of smoothness and feature representation. On one hand, the Gaussian activation function is smooth and infinitely differentiable, which allows the network to fit data in a more nuanced manner. Additionally, the shape of the Gaussian function enables each activation to create local features in the input space with varying scales and positions, thereby allowing the network to make finer adjustments within specific input regions. In contrast, the ReLU activation function is non-smooth (discontinuous at zero) and its derivative is constant in the non-zero region. This results in a relatively lower frequency response capability for MLPs with ReLU activations, as their activation characteristics are limited and do not represent the function space as intricately as Gaussian functions. Therefore, under the same depth and width, Gaussian-MLPs generally exhibit a larger network bandwidth due to their greater smoothness and more complex feature representation capabilities. Consequently, for a given audio input (with \( N \) being constant), the bandwidth \( D \) of Gaussian-MLPs is greater than that of ReLU-MLPs, which aligns with our benchmark results indicating that Gaussian-MLPs are better suited for audio signal representation.

Furthermore, since positional encoding maps low-dimensional input coordinates into a higher-dimensional space, it enhances the network's bandwidth $D$. Therefore, the embedding of positional encodings can effectively improve the capability of Coordinate-MLPs to represent audio signals.

\subsection{B.2 Memorization of Fourier-KAN}
Liu et al. \cite{liu2024kan} demonstrated that Kolmogorov–Arnold networks employing B-splines can approximate arbitrary functions with an extremely low error margin. We will outline the main proof process below.

\textbf{Derivation.}
A KAN network with $L$ layers can be expressed as,
\begin{equation}
    \label{eq:kan}
    f(\mathbf{x})=(\mathbf{\Phi}_{L-1}\circ \mathbf{\Phi}_{L-2} \cdots \mathbf{\Phi}_{1}  \circ \mathbf{\Phi}_{0})(\mathbf{x})
\end{equation} where $\Phi_l$ represents the transfer matrix between $(l-1)$-th layer and $l$-th layer,
$$
\mathbf{\Phi}_l={\left(\begin{array}{cccc}
\phi_{l, 1,1}(\cdot) & \phi_{l, 1,2}(\cdot) & \cdots & \phi_{l, 1, n_l}(\cdot) \\
\phi_{l, 2,1}(\cdot) & \phi_{l, 2,2}(\cdot) & \cdots & \phi_{l, 2, n_l}(\cdot) \\
\vdots & \vdots & & \vdots \\
\phi_{l, n_{l+1}, 1}(\cdot) & \phi_{l, n_{l+1}, 2}(\cdot) & \cdots & \phi_{l, n_{l+1}, n_l}(\cdot)
\end{array}\right)}.
$$

According to the classical theory of one-dimensional B-splines \cite{Boor1978APG}, and given that $\Phi_{l, i, j}$ are continuous functions that can be uniformly bounded on a bounded domain, it follows that there exist finite-grid B-spline functions $\Phi_{l, i, j}^G$ such that for any $0 \leq m \leq k$,
\begin{equation*}
    \begin{aligned}
        & \| \left(\Phi_{l, i, j} \circ \boldsymbol{\Phi}_{l-1} \circ \boldsymbol{\Phi}_{l-2} \circ \cdots \circ \boldsymbol{\Phi}_1 \circ \boldsymbol{\Phi}_0\right) \mathbf{x} - \\
        & \left({\Phi}_{l, i, j}^G \circ \boldsymbol{\Phi}_{l-1} \circ \boldsymbol{\Phi}_{l-2} \circ \cdots \circ \boldsymbol{\Phi}_1 \circ \boldsymbol{\Phi}_0\right) \mathbf{x}{\|}_{C^m} \leq C^{-k-1+m},
    \end{aligned}
\end{equation*}
with a constant $C$ independent of $G$, and the $C^m$-norm is utilized to measure the magnitude of derivatives up to order $m$,
$$
 \|g\|_{C^m}=\max _{|\beta| \leq m} \sup _{x \in[0,1]^n}\left|D^\beta g(x)\right|.
$$
If we fix these B-spline approximations, the residual $R_l$ can be defined as follows,
\begin{equation*}
    \begin{aligned}
        R_l:=&\left(\boldsymbol{\Phi}_{L-1}^G \circ \cdots \circ \boldsymbol{\Phi}_{l+1}^G \circ \boldsymbol{\Phi}_l \circ \boldsymbol{\Phi}_{l-1} \circ \cdots \circ \boldsymbol{\Phi}_0\right) \mathbf{x}-\\
        &\left(\boldsymbol{\Phi}_{L-1}^G \circ \cdots \circ \boldsymbol{\Phi}_{l+1}^G \circ \boldsymbol{\Phi}_l^G \circ \boldsymbol{\Phi}_{l-1} \circ \cdots \circ \boldsymbol{\Phi}_0\right) \mathbf{x},
    \end{aligned}
\end{equation*}
which satisfies,
$$
\left\|R_l\right\|_{C^m} \leq C G^{-k-1+m},
$$
with a constant independent of $G$. Finally notice that
\begin{equation*}
    \begin{aligned}
        f-&\left(\boldsymbol{\Phi}_{L-1}^G \circ \boldsymbol{\Phi}_{L-2}^G \circ \cdots \circ \boldsymbol{\Phi}_1^G \circ \boldsymbol{\Phi}_0^G\right) \mathbf{x}=\\
        &R_{L-1}+R_{L-2}+\cdots+R_1+R_0.
    \end{aligned}
\end{equation*}

Based on the above derivation, we can conclude the following. For the function \( f(\mathbf{x}) \) in Eq. \ref{eq:kan}, if each \( \Phi_{l, i, j} \) is \((k+1)\)-times continuously differentiable, then there exists a constant \( C \), dependent on \( f \) and its representation, such that the following approximation bound holds with respect to the grid size \( G \): there exist \( k \)-th order B-spline functions \( \Phi_{l, i, j}^G \) such that for any \( 0 \leq m \leq k \), the following inequality is satisfied:
$$
\left\|f-\left(\boldsymbol{\Phi}_{L-1}^G \circ \boldsymbol{\Phi}_{L-2}^G \circ \cdots \circ \boldsymbol{\Phi}_1^G \circ \boldsymbol{\Phi}_0^G\right) \mathbf{x}\right\|_{C^m} \leq C G^{-k-1+m}.
$$

\textbf{Analysis.}
Liu et al. \cite{liu2024kan} have demonstrated that KAN networks based on B-spline basis functions are capable of approximating arbitrary functions within a specified error margin. Consequently, if it can be shown that B-spline basis functions can be approximated by Fourier basis functions, it follows that Fourier-KAN networks can also approximate any signal function within a given error tolerance. We will outline the main proof process below.

Given a B-spline function \( S(x) \) defined over the interval \([a, b]\),
\[
S(x) = \sum_{i=0}^{n} c_i B_{i,k}(x),
\]
where \( c_i \) are the corresponding control points, and \( B_{i,k}(x) \) represents the \(k\)-th order B-spline basis functions, which is defined through the following recursive relations,
\begin{equation*}
\begin{aligned}
        B_{i,0}(x) &= \begin{cases}
       1 & \text{if } t_i \leq x < t_{i+1}, \\
       0 & \text{otherwise},
   \end{cases} \\
     B_{i,k}(x) &= \frac{x - t_i}{t_{i+k} - t_i} B_{i,k-1}(x) + \frac{t_{i+k+1} - x}{t_{i+k+1} - t_{i+1}} B_{i+1,k-1}(x),
\end{aligned}
\end{equation*} where \( t_i \) represents the sequence of knots, and \( k \) denotes the order of the spline. This recursive formula constructs higher-order B-spline basis functions \( B_{i,k}(x) \) from lower-order basis functions \( B_{i,k-1}(x) \) and \( B_{i+1,k-1}(x) \).

To transform the B-spline function into a periodic function, we define the periodic extension function \( \tilde{S}(x) \),
\[
\tilde{S}(x) = \begin{cases}
S(x) & \text{if } x \in [a, b], \\
S(x - T) & \text{if } x \in [a + T, b + T],
\end{cases}
\] where \( T = b - a \) represents the period. This extension process converts the non-periodic B-spline function into a periodic function.

Consider the periodic extension \( \tilde{S}(x) \). According to the properties of Fourier series, any periodic function \( f(x) \) can be expressed as,
\[
f(x) = a_0 + \sum_{n=1}^{\infty} \left[ a_n \cos\left(\frac{2\pi n x}{T}\right) + b_n \sin\left(\frac{2\pi n x}{T}\right) \right],
\]
where the Fourier coefficients \( a_n \) and \( b_n \) are computed as follows,
\[
a_0 = \frac{1}{T} \int_{0}^{T} f(x) \, dx,
\]
\[
a_n = \frac{2}{T} \int_{0}^{T} f(x) \cos\left(\frac{2\pi n x}{T}\right) \, dx,
\]
\[
b_n = \frac{2}{T} \int_{0}^{T} f(x) \sin\left(\frac{2\pi n x}{T}\right) \, dx.
\]

According to the Fourier series approximation theorem \cite{Davidson2010}, any square-integrable periodic function \( f(x) \) (including \( \tilde{S}(x) \)) can be approximated by its Fourier series in the \( L^2 \) sense, with the approximation error decreasing as the number of terms \( N \) increases,
\[
\| \tilde{S}(x) - S_N(x) \|_{L^2} \to 0 \text{ as } N \to \infty,
\]
where \( S_N(x) \) represents the partial sum of the Fourier series with \( N \) terms:
\[
S_N(x) = a_0 + \sum_{n=1}^{N} \left[ a_n \cos\left(\frac{2\pi n x}{T}\right) + b_n \sin\left(\frac{2\pi n x}{T}\right) \right].
\]

In summary, since a B-spline function \( S(x) \) is a piecewise polynomial defined over a finite interval, its periodic extension \( \tilde{S}(x) \) is a periodic function. Furthermore, \( \tilde{S}(x) \) can be approximated in the \( L^2 \) space by its Fourier series in the mean square sense. Therefore, any B-spline function \( S(x) \) can be approximated by its Fourier series \( S_N(x) \) through its periodic extension \( \tilde{S}(x) \), with the approximation error approaching zero as the number of Fourier series terms \( N \) increases.

\textbf{Conclusion.}
Based on the above analysis, we indirectly demonstrate that Fourier-KAN networks can approximate signals of arbitrary complexity. As the complexity of the signal changes, we can dynamically adjust the frequency grid size \(\Omega\) (see ``Frequency-adaptive Learning Strategy'' section in our paper). This adjustment reflects the ability to capture different frequency components of the signal and enhance the representational capacity of the Fourier-KAN network. Additionally, only a sufficiently large frequency grid size can approximate complex audio signals, which provides a theoretical foundation for the use of an inverted pyramid frequency setting.

\subsection{B.3 Parameter Utilization Comparison}
As demonstrated in Sections B.1 and B.2, both Coordinate-MLPs and Fourier-KAN theoretically possess the capability to memorize signals of arbitrary complexity. However, from the perspective of nonlinearity, it is evident that Fourier-KAN exhibits superior representational power. Nevertheless, assessing the memory capacity of these models remains challenging. Therefore, we conduct a comparative analysis of the storage capacity of Coordinate-MLPs and Fourier-KAN based on the utilization of their parameters.

\textbf{Analysis.} 
For simplicity, consider a network with depth \(L\), where each layer has a uniform width \(n_0 = n_1 = \cdots = n_L = N\) and the maximum frequency threshold is consistent across layers, \(\Omega_1 = \Omega_2 = \cdots = \Omega_L = \Omega\). The total number of parameters in such a Fourier-KAN is given by,
\[
P_{\text{Fourier-KAN}} = O(N^2 L\Omega).
\]
In contrast, the parameter count for a multilayer perceptron (MLP) with the same depth \(L\) and width \(N\) is,
\[
P_{\text{MLP}} = O(N^2 L)
\].
This implies that when representing discrete signals of equivalent capacity, an MLP typically requires more layers, which may lead to issues such as gradient explosion or vanishing gradients. Therefore, the parameter efficiency of the Fourier-KAN is significantly superior to that of the MLP.

\begin{table}[!htbp]
  \centering
  \tabcolsep=1.0mm
   \fontsize{9}{11}\selectfont
  \begin{tabular}{ccccc}
    \cline{1-5}
     Model &  Network & Params & SNR $\uparrow$ & LSD $\downarrow$ \\
     
    \cline{1-5}
    ReLU & [1,256,256,256,256,1] & \underline{ 263K} & 0.00 & 4.623 \\

    \cline{1-5}
    RFF+ReLU & [1,256,256,256,256,1] & 280K & 15.62 & \underline{ 0.978} \\

     \cline{1-5}
    NeFF+ReLU & [1,256,256,256,256,1] & 270K & \underline{ 22.29} & 1.129 \\

      \cline{1-5}
    Fourier-KAN & [1,64,64,64,64,1] & \fs 254K & \fs 33.14 & \fs 0.961 \\
  \cline{1-5}
  \end{tabular}
 \caption{Comparison of parameter utilization between Fourier-KAN and ReLU-MLPs on ``Bach'' audio clips.}
  \label{tab:parameter_usage}
\end{table}

\textbf{Experiments. } 
We compare the parameter counts and the capacity for representing audio signals between ReLU-MLPs and Fourier-KAN networks. The ReLU-MLPs under consideration have a network depth of 6 and a width of 256. In contrast, the Fourier-KAN network has a depth of 6 and a width of 64, with maximum frequency thresholds \(\Omega\) set to 1024, 5, and 3 for the input, hidden, and output layers, respectively.

As shown in Table \ref{tab:parameter_usage}, the parameter count of a Fourier-KAN with a width of only 64 is similar to that of ReLU-MLPs with a width of 256 when the depth is kept constant. This indicates that Fourier-KAN does not require complex networks to represent audio signals of the same capacity. Furthermore, due to the periodic nature of the Fourier basis functions used in Fourier-KAN, it can represent audio signals more effectively with fewer parameters than ReLU-MLPs, demonstrating higher parameter efficiency.

\section{C. Interpretability}
Completely explaining neural networks using symbolic regression is challenging and often futile. However, it is possible to perform interpretability analysis of the representational form of the network from a local perspective. Our Fourier-KAN is derived from the assumption of local periodicity and the Fourier series theorem. Consequently, we will briefly analyze the interpretability of both Coordinate-MLPs and Fourier-KAN through these two concepts.

\begin{table}[!ht]
    \centering
    \resizebox{0.48\textwidth}{!}{
    \begin{tabular}{|c|}
        \hline
        \begin{minipage}[t]{\linewidth}
            \begin{assumption}\label{asp:lpa_appendix}
                \textbf{Local Periodicity Assumption.} 
                For a complex non-stationary signal \( f(t) \), there exists a sufficiently small time interval \( \epsilon > 0 \) where \( f(t) \) can be approximated by a periodic function \( p(t) \):
                $$
                \exists \epsilon > 0, \quad \forall t, \quad |t| < \epsilon: \quad f(t) \approx p(t)
                $$
                where \( p(t) \) is a periodic function with period \( T \), satisfying \( p(t + T) = p(t) \).
            \end{assumption}
        \end{minipage} \\ 
        \hdashline
        \begin{minipage}[t]{\linewidth}
            \begin{theorem}\label{thm1_appendix}
               \textbf{Fourier Series Theorem.} 
                Any periodic signal \( p(t) \) with period \( T \) can be represented as an infinite series of sine and cosine functions,
                $$
                p(t) = \frac{a_0}{2} + \sum_{n=1}^{\infty} \left( a_n \cos\left(\frac{2\pi nt}{T}\right) + b_n \sin\left(\frac{2\pi nt}{T}\right) \right)
                $$
                where \( a_0 \), \( a_n \), and \( b_n \) are the Fourier coefficients.
            \end{theorem}
        \end{minipage} \\ 
        \hline
    \end{tabular}
    }
\end{table}

According to the assumption of local periodicity (Assumption \ref{asp:lpa_appendix}) and the Fourier series theorem (Theorem \ref{thm1_appendix}), any complex audio signal \( f(t) \) can be regarded as periodic within a \textbf{local} region. Consequently, it can be expressed as a series of sine and cosine functions,
\begin{equation}
    \label{eq:local_fourier}
    f(t) = \frac{a_0}{2} + \sum_{n=1}^{\infty} \left( a_n \cos\left(\frac{2\pi nt}{T}\right) + b_n \sin\left(\frac{2\pi nt}{T}\right) \right).
\end{equation}

\subsection{C.1 Interpretability of Coordinate-MLPs}
\textbf{Analysis.} For most activation functions, such as ReLU, Tanh, and Gaussian, the learning ability of the network can primarily be explained from the perspective of approximating the function \( f(t) \) (Eq. \ref{eq:local_fourier}). Consequently, the smoothness, continuity, and differentiability of these activation functions may impact the network's capacity to represent audio signals. For example, the Gaussian activation function is smoother and more continuous than ReLU, which explains why Gaussian-MLPs have a superior ability to fit audio signals compared to ReLU-MLPs. This finding is consistent with the conclusion we reached in Sect. 
 B.1\ref{memorization_mlps}.

In the case of activation functions such as Sine, their interpretability can be approached from the perspective of the Fourier series theorem.

Consider the input layer of Sine-MLPs: let the input time coordinate be denoted as \(t \in \mathbb{R}\), and suppose the weight matrix is \(\mathbf{W} \in \mathbb{R}^{3\times 1}\) and the bias vector is \(\mathbf{b} \in \mathbb{R}^{3\times 1}\). Then, the output of the input layer, \(\mathbf{y} \in \mathbb{R}^{3\times 1}\), can be expressed as follows,
$$
\mathbf{y} = \sin(\omega \mathbf{W} t + \mathbf{b})
$$ 
The expansion of this expression is given by, 
$$
\mathbf{y} = \sin(\omega \mathbf{W} t + \mathbf{b}) = \begin{bmatrix} \sin(\omega w_1 t + b_1) \\ \sin(\omega w_2 t + b_2) \\ \sin(\omega w_3 t + b_3) \end{bmatrix},
$$then,
\begin{equation}
\label{eq:sine_input}
    \begin{aligned}
        y_1 &= \sin(\omega w_1 t + b_1) \\
        y_2 &= \sin(\omega w_2 t + b_2) \\
        y_3 &= \sin(\omega w_3 t + b_3). 
    \end{aligned}
\end{equation}
This implies that the primary function of the input layer is to map the time coordinates into a higher-dimensional space, akin to the role of Fourier feature positional encoding (e.g., \texttt{NeFF}, \texttt{RFF}). Consequently, in applications such as radiance fields and image representation, Sine-type activation functions can effectively eliminate the need for positional encoding.

Considering the output layer of Sine-MLPs: 
\begin{itemize}
    \item Input hidden variables:  \(\mathbf{z} \in \mathbb{R}^3\), where \(\mathbf{z} = [z_1, z_2, z_3]^T \).
    \item Bias vector: \(\mathbf{c} \in \mathbb{R}^3\), where \(\mathbf{c} = [c_1, c_2, c_3]^T\).
    \item Weight vector: \(\mathbf{v} \in \mathbb{R}^3\), where \(\mathbf{v} =[v_1, v_2, v_3]^T\).
    \item Frequency scalar: \(\omega \in \mathbb{R}\).
    \item Output amplitude: \(A \in \mathbb{R}\).
\end{itemize}
Then the output of the Sine-MLP's output layer in matrix form can be expressed as,
$$
A = \sin(\omega \cdot (\mathbf{v}^T \mathbf{z} + \mathbf{c}^T \mathbf{1}))
$$ Where \(\mathbf{1}\) is a vector of ones \([1, 1, 1]^T\), which ensures that the bias is correctly added to the linear transformation. Expanding this expression, we can have,
\begin{equation}
\label{eq:sine_output}
    A = \sin(\omega \cdot (v_1 z_1 + v_2 z_2 + v_3 z_3 + c_1 + c_2 + c_3)).
\end{equation}

\textbf{Conclusion.}
For the majority of activation functions (ReLU, Gaussian, etc.), evaluation is primarily based on their ability to approximate audio signals, considering properties such as smoothness, continuity, and differentiability. In contrast, for periodic Sine-type activation functions, as shown in Eq. \ref{eq:sine_input}, the input layer effectively functions as Fourier feature positional encoding. This explains why such functions exhibit strong representational capability in applications such as radiance fields and image representation, even without explicit positional encoding. On the other hand, as indicated in Eq \ref{eq:sine_output}, the output layer represents audio amplitudes using a sine function with a single frequency \(\omega\). This representation fundamentally involves feature fusion, albeit with periodic characteristics localized to the network. Given that different audio signals exhibit varying frequency scales, this also elucidates why Sine-type activation functions are sensitive to frequency hyperparameters and dependent on frequency-related initialization schemes.

\subsection{C.2 Interpretability of Fourier-KAN}
\textbf{Analysis.} Given a Fourier-KAN with dimensions \([n_0, n_1, \cdots, n_L]\), it can be defined using the following recursive relationship,
\begin{equation*}
\begin{aligned}
    \mathbf{z}^{(0)} & = {t} \\
    \mathbf{z}^{(l+1)} &= \sum_{\omega=1}^{\Omega_l} \left[ \mathbf{a}^{(l,\omega)}\cos(\omega \mathbf{z}^{(l)}) + \mathbf{b}^{(l,\omega)}\sin(\omega \mathbf{z}^{(l)})\right] + \mathbf{c}^{(l)} \\
    f({t}) & = \mathbf{z}^{(n_L)}\left(\mathbf{z}^{(n_L-1)}\left(\mathbf{z}^{(l)}\left(...(\mathbf{z}^{(0)})\right)\right)\right), l=n_0, \ldots, n_L, \\
\end{aligned}
\end{equation*}  where \(\mathbf{a}^{(l, \omega)}\) and \(\mathbf{b}^{(l, \omega)} \in \mathbb{R}^{d_{l+1} \times d_l}\) represent the Fourier coefficient weights for the \(l\)-th layer at frequency \(\omega\) respectively, \(\mathbf{c}^{(l)} \in \mathbb{R}^{d_{l+1}}\) is the bias term of $l$-th layer, \(z^{(l)} \in \mathbb{R}^{d_l}\) denotes the hidden units of the \(l\)-th layer, and \(\Omega_l\) is a hyperparameter indicating the maximum frequency threshold for the \(l\)-th layer.

Considering the input layer of a Fourier-KAN:
\begin{itemize}
    \item Input time coordinate: $t \in \mathbb{R}$.
    \item Output dimension: $d_1=3$.
    \item Frequency \(\Omega_0\) for the first layer is 3.
    \item  \(\mathbf{a}^{(0,\omega)} = [a^{(0,\omega)}_1, a^{(0,\omega)}_2, a^{(0,\omega)}_3]^T\) denotes the weight coefficients associated with the cosine terms for frequency \(\omega\).
    \item \(\mathbf{b}^{(0,\omega)} = [b^{(0,\omega)}_1, b^{(0,\omega)}_2, b^{(0,\omega)}_3]^T\) denotes the weight coefficients associated with the sine terms for frequency \(\omega\).
    \item \(\mathbf{c}^{(0)} = [c^{(0)}_1, c^{(0)}_2, c^{(0)}_3]^T\) is the bias term vector for the first layer.
\end{itemize}
Given these definitions, the output of the first layer, denoted by \(\mathbf{z}^{(1)}\), can be expressed as,
\begin{equation*}
    \begin{aligned}
        \mathbf{z}^{(1)} &= \begin{bmatrix}
        \cos(t) & \sin(t) & \cos(2t) & \sin(2t) & \cos(3t) & \sin(3t)
        \end{bmatrix} \\
        &\quad \times \begin{bmatrix}
        a^{(0,1)}_1 \\
        b^{(0,1)}_1 \\
        a^{(0,2)}_1 \\
        b^{(0,2)}_1 \\
        a^{(0,3)}_1 \\
        b^{(0,3)}_1
        \end{bmatrix}
        + \begin{bmatrix}
        c^{(0)}_1 \\
        c^{(0)}_2 \\
        c^{(0)}_3
        \end{bmatrix}.
    \end{aligned}
\end{equation*}
In expanded form, the output of the first layer \(\mathbf{z}^{(1)}\) is:
\begin{equation}
\label{eq:fourier_input}
\begin{aligned}
    \mathbf{z}^{(1)}_1 &= a^{(0,1)}_1 \cos(t) + b^{(0,1)}_1 \sin(t) \\
    & + a^{(0,2)}_1 \cos(2t) + b^{(0,2)}_1 \sin(2t)\\
    & + a^{(0,3)}_1 \cos(3t) + b^{(0,3)}_1 \sin(3t) \\ 
    & + c^{(0)}_1 \\
    \mathbf{z}^{(1)}_2 &= a^{(0,1)}_2 \cos(t) + b^{(0,1)}_2 \sin(t) \\ &
    + a^{(0,2)}_2 \cos(2t) + b^{(0,2)}_2 \sin(2t) \\ &
    + a^{(0,3)}_2 \cos(3t) + b^{(0,3)}_2 \sin(3t)\\ &
    + c^{(0)}_2 \\
    \mathbf{z}^{(1)}_3 &= a^{(0,1)}_3 \cos(t) + b^{(0,1)}_3 \sin(t)\\
    & + a^{(0,2)}_3 \cos(2t) + b^{(0,2)}_3 \sin(2t)\\
    & + a^{(0,3)}_3 \cos(3t) + b^{(0,3)}_3 \sin(3t) \\
    &+ c^{(0)}_3 \\
\end{aligned}
\end{equation}
Comparing Eq. \ref{eq:local_fourier} with Eq. \ref{eq:fourier_input} reveals that the input layer of the Fourier-KAN maps the time coordinate \( t \) into a high-dimensional Fourier space, incorporating various frequencies. This allows the network to learn frequency components across different frequency domains.

We now turn our attention to the output layer of the Fourier-KAN:
\begin{itemize}
    \item Input to the output layer: \(\mathbf{z}^{(L)} \in \mathbb{R}^3\).
    \item  Output dimension: \(d_{L+1} = 1\).
    \item  Output amplitude: $A \in \mathbb{R}$.
    \item Frequency \(\Omega_L\) for the output layer is 3.
    \item \(\mathbf{a}^{(L,\omega)} = [a^{(L,\omega)}_1, a^{(L,\omega)}_2, a^{(L,\omega)}_3]^T\) represents the weight coefficients associated with the cosine terms for frequency \(\omega\) in the output layer.
    \item \(\mathbf{b}^{(L,\omega)} = [b^{(L,\omega)}_1, b^{(L,\omega)}_2, b^{(L,\omega)}_3]^T\) represents the weight coefficients associated with the sine terms for frequency \(\omega\) in the output layer.
    \item \(\mathbf{c}^{(L)} = [c^{(L)}]\) is the bias term for the output layer, which is a scalar.
\end{itemize} 
Then, the output amplitude of the output layer is given by,
\begin{equation*}
    \begin{aligned}
        A &= \begin{bmatrix}
        \cos(\mathbf{z}^{(L)}) & \sin(\mathbf{z}^{(L)}) & \cdots & \cos(3\mathbf{z}^{(L)}) & \sin(3\mathbf{z}^{(L)})
        \end{bmatrix} \\
        &\quad \times \begin{bmatrix}
        a^{(L,1)}_1 \\
        b^{(L,1)}_1 \\
        a^{(L,2)}_1 \\
        b^{(L,2)}_1 \\
        a^{(L,3)}_1 \\
        b^{(L,3)}_1
        \end{bmatrix}
        + \begin{bmatrix}
        c^{(L)}
        \end{bmatrix}.
    \end{aligned}
\end{equation*}
Expanding the above expression yields,
\begin{equation}
\label{eq:fourier_out}
\begin{aligned}
    A &= a^{(L,1)}_1 \cos(\mathbf{z}^{(L)}) + b^{(L,1)}_1 \sin(\mathbf{z}^{(L)}) \\
    &\quad + a^{(L,2)}_1 \cos(2\mathbf{z}^{(L)}) + b^{(L,2)}_1 \sin(2\mathbf{z}^{(L)}) \\
    &\quad + a^{(L,3)}_1 \cos(3\mathbf{z}^{(L)}) + b^{(L,3)}_1 \sin(3\mathbf{z}^{(L)}) \\
    &\quad + c^{(L)}.
\end{aligned}
\end{equation}
Comparing Eq.\ref{eq:local_fourier} with Eq. \ref{eq:fourier_out} reveals that the output layer of the Fourier-KAN inherently adopts the form of a Fourier series. This implies that the Fourier-KAN implicitly predicts the local neighborhood of the coordinate \( t \) using a Fourier series representation, which aligns with the assumption of local periodicity.

\textbf{Conclusion.} 
As analyzed above, the Fourier-KAN maps the time coordinate \( t \) into a high-dimensional space with different frequencies according to the Fourier series theorem, enabling the learning of features in neighborhoods with various frequencies. At the output layer, the Fourier-KAN represents the neighborhood features as corresponding amplitude values using a Fourier series. This indicates that the Fourier-KAN characterizes the neighborhood of the discrete coordinate \( t \) in the form of a Fourier series, thereby generating a continuous audio signal.

Compared to Coordinate-MLPs, the Fourier-KAN offers enhanced interpretability. We believe that the Fourier-KAN holds significant potential for further applications in the field of audio.

\section{D. Further Experiments}

\subsection{D.1 Experimental Setup}
\subsubsection{Evaluation Metrics.} 
Given a reference signal \( y \) and its approximation \( \hat{y} \), the Signal to Noise Ratio (SNR) is defined as follows \cite{roux2018snr},
$$
\operatorname{SNR}(\hat{y}, y)= 20 \log_{10} \left( \frac{\|y\|_2^2}{\|\hat{y}-y\|_2^2} \right).
$$
The SNR is a standard metric employed in the signal processing literature to quantify the quality of signal approximations. The Log-Spectral Distance (LSD), on the other hand, assesses the reconstruction quality across individual frequencies and is given by \cite{gray1976lsd},
$$
\operatorname{LSD}(\hat{y}, y) = \frac{1}{L} \sum_{\ell=1}^L \sqrt{\frac{1}{K} \sum_{k=1}^K (X(\ell, k) - \hat{X}(\ell, k))^2},
$$ where \( X \) and \( \hat{X} \) are the log-spectral power magnitudes of \( y \) and \( \hat{y} \), respectively, defined as \( X = \log |S|^2 \), with \( S \) being the short-time Fourier transform (STFT) of the signal. Here, \( \ell \) and \( k \) index the frames and frequencies, respectively. In our experiments, we utilized frames of length 2048. Since LSD is an indirect measure for evaluation in the frequency domain, we focus primarily on the SNR metric.

\subsubsection{Implementation Details.}
Our NeAF model was implemented in Python using the PyTorch framework and was trained on a workstation equipped with an AMD EPYC 7302 16-core processor and an NVIDIA GeForce RTX 3090 GPU. The training process spanned 1000 epochs with a batch size of 16,384, utilizing a learning rate of 1e-4 and a learning rate scheduling strategy based on the CosineAnnealingLR method. For the \texttt{RFF} positional encoding, the variance $\sigma$ was set to 10, and the frequency dimension $L$ was 32. Regarding the \texttt{NeFF} positional encoding, the frequency dimension was defined as $log(1 / (2 * (2 * 1 / B)))$, where \(B\) represents the batch size. The frequency parameter $\omega$ for the Sine and Incode-Sine activation functions was set to 30. Additionally, for the Gaussian-type activation functions, the variance factor $a$ was fixed at 0.1.

\subsubsection{Dataset Details.}
\textbf{GTZAN} \cite{tzanetakis2001gtzan} music dataset contains 1000 music snippets of ten different genres at 30 seconds each. \textbf{CSTR VCTK} \cite{yamagishi2019cstr} corpus includes speech data uttered by 110 English speakers with various accents. Each speaker read out about 400 sentences, which were selected from a newspaper, the rainbow passage and an elicitation paragraph used for the speech accent archive. 

For the GTZAN dataset \cite{tzanetakis2001gtzan}, we selected the audio file numbered 00000 from each of the ten music genres (blues, classical, country, disco, hip-hop, jazz, metal, pop, reggae, and rock) as the experimental audio data.

For the CSTR VCTK dataset \cite{yamagishi2019cstr}, we selected audio data from ten different accents (English, Scottish, Northern Irish, Irish, Indian, Welsh, New Zealand, American, Canadian, and Australian), which represent the longest recordings in the dataset. The corresponding speaker IDs are 225, 234, 238, 245, 248, 253, 335, 345, 363, and 374, with sequence number 023 and microphone number 1.

\subsection{D.2 Supplement to Benchmark Leaderboard}
As illustrated in Table \ref{tab:mlp_benchmark_appendix}, we provide a comprehensive benchmark of Coordinate-MLPs in audio signal representations and additionally include the evaluation results for two Kolmogorov–Arnold Networks: Bspline-KAN and our proposed Fourier-ASR.

From this comprehensive benchmark, we derive the following detailed conclusions,
\begin{itemize}
    \item The majority of activation functions (such as Gabor-Wavelet, Quadratic, Multi-Quadratic, ExpSin, Sigmoid, SoftPlus, Tanh, ELU, SiLU, PReLU, ReLU) are inadequate for representing audio signals without positional encoding.

    \item Only highly nonlinear functions (such as Gaussian, Laplacian, Super-Gaussian) or periodic functions (such as Sine, Incode-Sine) are capable of partially capturing the high-frequency and local periodic characteristics of audio signals.

    \item Positional encoding significantly enhances the capability of Coordinate-MLPs to represent audio signals. This improvement is primarily attributed to the high-dimensional mapping of positional encoding, which increases the model's nonlinearity and bandwidth (see appendix B.1).

    \item Due to the periodicity of Fourier-KAN, our proposed Fourier-ASR outperforms the B-Spline-based KAN network \cite{liu2024kan}.

    \item In the absence of positional encoding, our proposed Fourier-ASR achieves state-of-the-art performance, markedly surpassing the next-best method (Sine) with an SNR improvement of 10.42 dB.
\end{itemize}

\begin{table*}[!ht]
\centering
\tabcolsep=0.5mm
\fontsize{9}{11}\selectfont
    \begin{tabular}{c|c|c|c|c|c|c|c|c|c|c|c}
  
        \cline{1-12}
        \multirow{2}{*}{\makecell{Activation $\sigma(\cdot)$}} & \multirow{2}{*}{\makecell{Equation}} & \multirow{2}{*}{\makecell{Parameter}} & \multirow{2}{*}{\makecell{PE $\gamma(\cdot)$}} & \multicolumn{2}{c|}{Bach \texttt{(7s)}} & \multicolumn{2}{c|}{Counting \texttt{(7s)}} & \multicolumn{2}{c|}{Blues \texttt{(30s)}} & \multicolumn{2}{c}{\texttt{Avg.}} \\
        
        \cdashline{5-12}
        & & & & SNR $\uparrow$ & LSD $\downarrow$ & SNR $\uparrow$ & LSD $\downarrow$ & SNR $\uparrow$ & LSD $\downarrow$ & SNR $\uparrow$ & LSD $\downarrow$ \\

        \cline{1-12}
        \multirow{3}{*}{\makecell{$\operatorname{Gabor-Wavelet}$}} & \multirow{3}{*}{\makecell{
        $e^{j a x} \cdot e^{-\left|b x\right|^2}$}} & \multirow{3}{*}{\makecell{$a,b$ }}& \texttt{Identity} & - & - & - & - & - & - & - & - \\
        & & & \texttt{RFF} & - & - & - & - & - & - & - & - \\
        & & & \texttt{NeFF} & - & - & - & - & - & - & - & - \\
        
        \cline{1-12}
        \multirow{3}{*}{\makecell{$\operatorname{Quadratic}$}} & \multirow{3}{*}{\makecell{
        $\frac{1}{1+(a x)^2}$}} & \multirow{3}{*}{\makecell{$[a]$ }}& \texttt{Identity} & - & - & -3.37 & 5.182 & -1.553 & 7.573 & - & -  \\
        & & & \texttt{RFF} & -1.55 & 6.346 & -8.28 & 5.182 & -1.97 & 7.572 & -3.93 & 6.367 \\
        & & & \texttt{NeFF} & -5.69 & 6.344 & -1.97 & 5.182 & -2.59 & 7.572 & -3.42 &  6.366 \\
        
        \cline{1-12}
        \multirow{3}{*}{\makecell{$\operatorname{Multi-Quadratic }$}} & \multirow{3}{*}{\makecell{
        $\frac{1}{\sqrt{1+(a x)^2}}$}} & \multirow{3}{*}{\makecell{$[a]$ }}& \texttt{Identity} & - & -2.48  & 5.182 & - & -1.55 & 7.573 & - & - \\
        & & & \texttt{RFF} & - & - & -1.19 & 5.182 & -3.11 & 7.573 & - & -\\
        & & & \texttt{NeFF} & -5.17 & 6.344 & -2.17 & 5.182 & -1.55 & 7.573 & -2.96 & 6.366 \\
        
        \cline{1-12}
        \multirow{3}{*}{\makecell{$\operatorname{ExpSin}$}} & \multirow{3}{*}{\makecell{
        $e^{-\sin (a x)}$}} & \multirow{3}{*}{\makecell{$[a]$}}& \texttt{Identity} & 0.00 & 6.339 & 0.00 & 5.180 & -8.28 & 7.571 & -2.76 & 6.363 \\
        & & & \texttt{RFF} & 7.97 & 5.297 & 1.45 & 4.171 & 0.00 & 6.178 & 3.14 & 5.215 \\
        & & & \texttt{NeFF} & 0.00 & 3.326 & 1.04 & 4.550 & 0.00 & 5.520 & 0.35 & 4.465 \\
        
        \cline{1-12}
        \multirow{3}{*}{\makecell{$\operatorname{Sigmoid}$}} & \multirow{3}{*}{\makecell{
        $\frac{1}{1+e^{-x}}$}} & \multirow{3}{*}{}& \texttt{Identity} & 0.00 & 6.339 & 0.00 & 5.181 & -6.73 & 7.572 & -2.24 & 6.364 \\
        & & & \texttt{RFF} & 0.03 & 3.186 & 7.04 & 3.945 & 0.00 & 5.921 & 2.36 & 4.351 \\
        & & & \texttt{NeFF} & 0.51 & 1.905 & 0.00 & 4.259 & 0.00 & 5.212 & 0.17 & 3.792 \\
        
        \cline{1-12}
        \multirow{3}{*}{\makecell{$\operatorname{SoftPlus}$}} & \multirow{3}{*}{\makecell{
        $\frac{1}{a} \log \left(1+e^{a x}\right)$}} & \multirow{3}{*}{\makecell{$[a]$ }}& \texttt{Identity} & -1.92 & 6.311 & 0.00 & 5.165 & -3.11 & 7.554 & -1.68 & 6.343 \\
        & & & \texttt{RFF} & 0.55 & 1.647 & 0.03 & 2.709 & 0.06 & 3.773 & 0.21 & 2.710 \\
        & & & \texttt{NeFF} & 3.04 & 1.469 & 0.26 & 1.933 & 0.51 & 2.724 & 1.27 & 2.042 \\
        
        \cline{1-12}
        \multirow{3}{*}{\makecell{$\operatorname{Tanh}$}} & \multirow{3}{*}{\makecell{$\frac{e^x - e^{-x}}{e^x + e^{-x}}$}} & \multirow{3}{*}{} & \texttt{Identity} & -7.25 & 6.334 & 0.00 & 5.181 & -3.11 & 7.574 & -3.45 & 6.363 \\
        & & & \texttt{RFF} & 7.95 & 1.082 & 0.97 & 1.786 & 1.32 & 2.420 & 3.41 & 1.763 \\
        & & & \texttt{NeFF} & 11.09 & 1.046 & 6.87 & 1.593 & 3.60 & 1.927 & 7.19 & 1.522 \\
        
        \cline{1-12}
        \multirow{3}{*}{\makecell{$\operatorname{ELU}$}} & \multirow{3}{*}{\makecell{$\begin{cases} 
        x, & \text{if } x > 0 \\ 
        ae^x - a, & \text{otherwise} 
        \end{cases}$}} & \multirow{3}{*}{\makecell{$[a]$}} & \texttt{Identity} & 0.00 & 5.979 & 0.00 & 5.104 & -4.40 & 7.535 & -1.47 & 6.206 \\
        & & &  \texttt{RFF} & 7.77 & 1.111 & 0.67 & 1.956 & 0.75 & 2.723 & 3.06 & 1.930 \\
        & & &  \texttt{NeFF} & 12.26 & 0.997 & 4.89 & 1.557 & 2.52 & 2.057 & 6.56 & 1.537 \\
        
        \cline{1-12}
        \multirow{3}{*}{\makecell{$\operatorname{SiLU}$}} & \multirow{3}{*}{\makecell{$\frac{x}{1+e^{-x}}$}} & \multirow{3}{*}{} & \texttt{Identity} & -5.18 & 6.265 & -6.21 & 5.174 & -2.07 & 7.560 & -4.49 & 6.333 \\
        & & & \texttt{RFF} & 7.51 & 1.104 & 0.51 & 1.952 & 0.74 & 2.643 & 2.92 & 1.900 \\
        & & & \texttt{NeFF} & 12.71 &  0.950 & 7.50 & 1.618 & 4.23 & 1.881 & 8.15 & 1.483 \\
    
        \cline{1-12}
        \multirow{3}{*}{\makecell{$\operatorname{PReLU}$}} & \multirow{3}{*}{\makecell{\begin{tabular}{l} 
        $\begin{cases}x, & \text{ if } x>0 \\ ax, & \text{otherwise} \end{cases}$
        \end{tabular}}} & \multirow{3}{*}{\makecell{$[a]$ }}& \texttt{Identity} & 0.00 & 4.724 & 0.00 & 4.630 & 0.00 & 7.031 & 0.00 & 5.462 \\
        & & & \texttt{RFF} & 13.42 & 1.010 & 3.38 & 1.437 & 2.50 & 2.035 & 6.43 & 1.494  \\
        & & & \texttt{NeFF} & 17.50 & 1.133 & 7.88 & 1.575 & 5.20 & 1.539 & 10.19 & 1.416 \\
        
        \cline{1-12}
        \multirow{3}{*}{\makecell{$\operatorname{ReLU}$}} & \multirow{3}{*}{\makecell{$\max(0, x)$}} & & \texttt{Identity} & 0.00 & 4.623 & -7.66 & 4.546 & 0.00 & 6.774 & -2.55 & 5.314 \\
        & & &  \texttt{RFF} & 15.62 &  0.978 & 4.93 & \fs 1.400 & 3.23 & 1.862 & 7.93 &  1.413 \\
        & & &  \texttt{NeFF} &  22.29 & 1.129 &  9.57 &  1.538 & 7.64 & 1.324 &  13.17 &  1.330 \\
        
        \cline{1-12}
        \multirow{3}{*}{\makecell{$\operatorname{Gaussian}$}} & \multirow{3}{*}{\makecell{$e^{\frac{-x^2}{2a^2}}$ }} & \multirow{3}{*}{ \makecell{$[a]$ }} & \texttt{Identity} & 6.35 & 1.130 & 0.74 & 2.165 &  0.68 &  3.059 & 2.59 & 2.118 \\
        & & & \texttt{RFF} & 20.85 & 2.046 & 12.14 & 3.195 &  11.80 & 1.346 & 14.93 & 2.196 \\
        & & & \texttt{NeFF} & 19.68 & 2.127 & 9.20 & 3.438 & 7.74 & 1.597 & 12.21 & 2.387 \\
        
       \hline
        \multirow{3}{*}{\makecell{$\operatorname{Laplacian}$}} & \multirow{3}{*}{\makecell{
        $e^{\frac{-|x|}{a}}$}} & \multirow{3}{*}{\makecell{$[a]$ }} & \texttt{Identity} &  12.04 &  0.932 &  1.34  & \fs 1.561 & \underline{ 1.37} & \underline{ 2.434} &  4.92 &  \underline{1.642} \\
        & & & \texttt{RFF} & 15.57 & 2.386 & 10.97 & 2.632 &  14.74 & \fs 1.112 & 13.76 & 2.043 \\
        & & & \texttt{NeFF} & 15.26 & 2.434 & 8.67 & 3.191 &  8.16 &  1.526 & 10.70 & 2.384 \\
        
        \cline{1-12}
        \multirow{3}{*}{\makecell{Super-Gaussian}} & \multirow{3}{*}{\makecell{
        $[e^{\frac{-x^2}{2a^2}}]^b$ }} & \multirow{3}{*}{\makecell{$[a,b]$ }}& \texttt{Identity} & 6.38 & 1.101 & 0.75 & 2.166 &  0.68 &  3.059 & 2.60 & 2.109 \\
        & & & \texttt{RFF} &  20.86 & 2.044 &  12.44 & 3.169 &  11.80 & 1.346 &  15.03 & 2.186 \\
        & & & \texttt{NeFF} & 19.69 & 2.127 & 9.20 & 3.437 & 7.74 & 1.597 & 12.21 & 2.387 \\

        \hline
        \multirow{3}{*}{\makecell{Sine}} & \multirow{3}{*}{\makecell{
        $\sin (\omega x)$}} & \multirow{3}{*}{\makecell{$[\omega]$}}& \texttt{Identity} & \underline{ 13.36} & \underline{ 0.838} & \underline{ 7.96} &  1.660 & \fs 7.47 & \fs 1.722 & \fs 9.59 & \fs 1.407 \\
        & & & \texttt{RFF} & \fs 39.02 & \fs 0.582 & \fs 13.06 & \underline{ 1.412} & \fs 16.57 & \underline{ 1.156} & \fs 22.88 & \fs 1.050 \\
        & & & \texttt{NeFF} & \fs 42.39 & \fs 0.537 & \fs 33.58 & \fs 0.914 & \fs 22.02 & \fs 0.696 & \fs 32.66 &  \fs 0.716 \\
        
        \hline
        \multirow{3}{*}{\makecell{Incode-Sine}} & \multirow{3}{*}{\makecell{
        $a\sin (b \omega x + c) + d$}} & \multirow{3}{*}{\makecell{$[\omega], a,b,c,d$ }}& \texttt{Identity} & \fs 15.98 & \fs 0.778 & \fs 8.16 & \underline{ 1.611} & 0.01 & 3.865 & \underline{ 8.05} &   2.085 \\
        & & & \texttt{RFF} & \underline{ 38.10} & \underline{ 0.595} & \underline{ 12.86} &  1.559 & \underline{ 15.13} &  1.241 & \underline{ 22.03} &  \underline{ 1.132} \\
        & & & \texttt{NeFF} & \underline{ 41.40} & \underline{ 0.556} & \underline{ 32.24} & \underline{ 1.038} & \underline{ 21.33} & \underline{ 0.763} & \underline{ 31.99} & \underline{ 0.786} \\

        \hline
        \hline
        Basis $\phi(\cdot)$ & Equation & Parameter & PE $\gamma(\cdot)$ & SNR $\uparrow$ & LSD $\downarrow$ & SNR $\uparrow$ & LSD $\downarrow$ & SNR $\uparrow$ & LSD $\downarrow$ & SNR $\uparrow$ & LSD $\downarrow$ \\
        
        \cline{1-12}
        \multirow{3}{*}{\makecell{Bspline-KAN}} & \multirow{3}{*}{\makecell{
        $\frac{ax}{1+e^{-x}} + spline(x)$}}
        & \multirow{3}{*}{\makecell{$a,$ \\ \texttt{spline-coefs} }} & \texttt{No} & 1.54 & 1.531  & 0.00  & 3.293 & 0.00 & 4.643 & 0.51 & 3.156 \\
        & & & \texttt{RFF} & - & - & - & - & - & - & - & - \\
        & & & \texttt{NeFF} & - & - & - & - & - & - & - & -\\
        
        \cline{1-12}
        \multirow{3}{*}{\makecell{ Fourier-ASR}} & \multirow{3}{*}{\makecell{
        $a\sin (\omega x)+b\cos(\omega x)$}}
        & \multirow{3}{*}{\makecell{$a, b$}} & \texttt{No} & \fs 33.14 & 0.961  & \fs 13.10  & 1.377 & \fs 13.80 & 1.240 & \fs 20.01 & 1.193 \\
        & & & \texttt{RFF} & 16.37 & 2.352 & \fs 18.36 & 2.625 & 12.59 & 1.305 &  15.77 & 2.094 \\
        & & & \texttt{NeFF} & 15.99 & 2.380 & 6.96 & 3.603 &  8.32 & 1.564 & 10.42 & 2.516 \\
         
         \cline{1-12}
    \end{tabular}
    \caption{Benchmark leaderboard. For different types of positional encodings (\texttt{Identity}, \texttt{RFF}, \texttt{NeFF}), best results are highlighted as  \textbf{first} and \underline{second}. Note that ``\(a\)'' denotes a learnable parameter, while ``\([a]\)'' denotes a hyperparameter, and ``-'' means the model fails to converge. }
    \label{tab:mlp_benchmark_appendix}
\end{table*}

\subsection{D.3 Qualitative Experiments}
\begin{figure*}[!ht]
    \centering
    \includegraphics[width=0.9\textwidth]{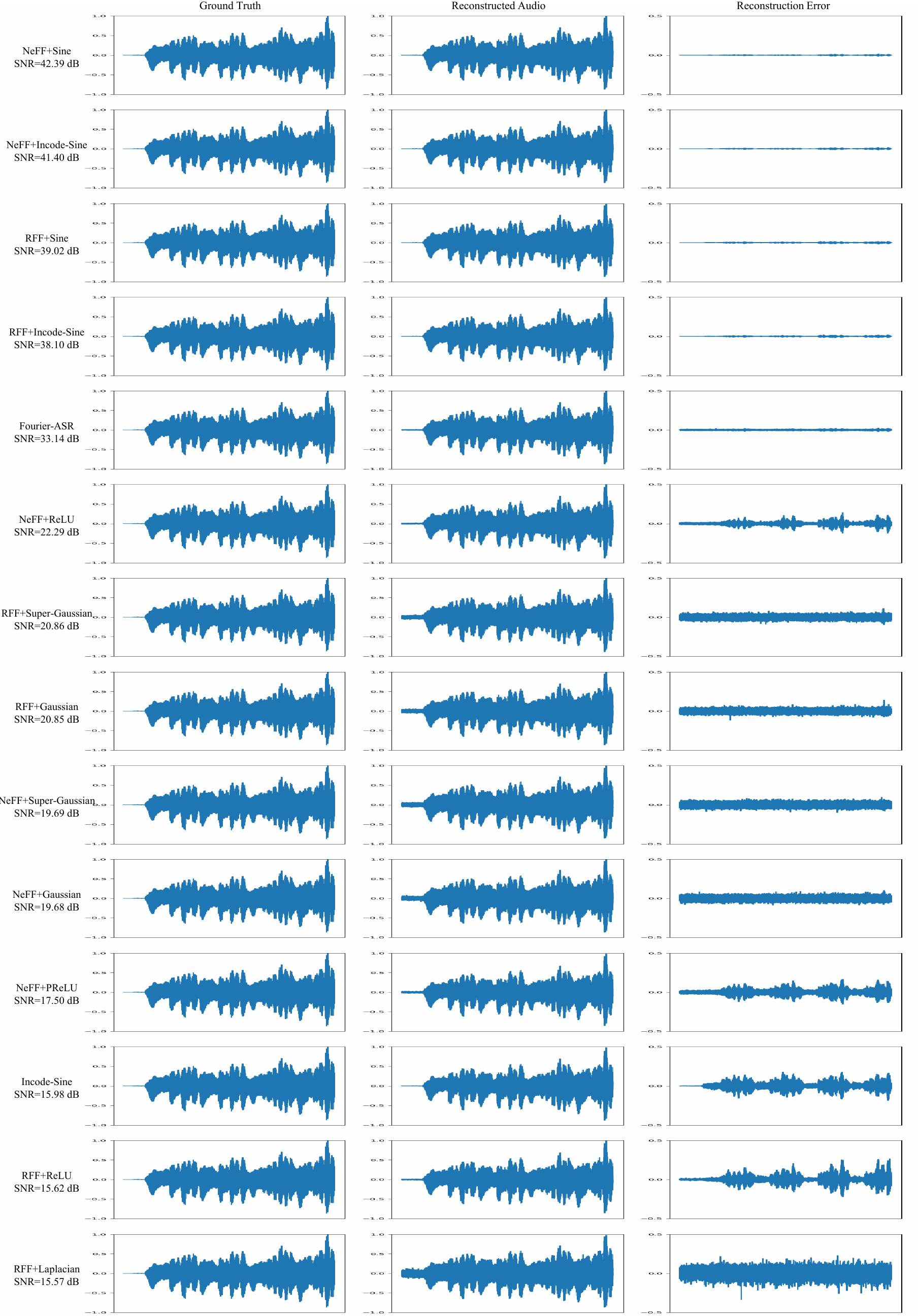}
    \caption{Qualitative experiments on ``Bach''. }
    \label{fig:audio_fig}
\end{figure*}
As illustrated in Fig. \ref{fig:audio_fig}, we conducted a qualitative experiment to provide a comprehensive benchmarking leaderboard. It is evident that the periodic Sine and Incode-Sine functions, when embedded with positional encoding, effectively capture high-frequency details of audio signals. Similarly, our proposed Fourier-ASR, which also exhibits periodicity, accurately represents audio signals even without the need for positional encoding. For highly nonlinear Gaussian activation functions, although their ability to represent audio signals is somewhat limited, their reconstruction error remains uniformly distributed across each time coordinate, unaffected by the amplitude of the audio. In contrast, the reconstruction errors for the periodic activation functions Sine and Incode-Sine vary with amplitude and are more pronounced in regions with significant changes, indicating a challenge in capturing the instantaneous variations in audio signals. Fourier-ASR, however, combines periodicity with strong nonlinearity, enabling it to capture instantaneous changes in audio more effectively, resulting in reconstructions that align more closely with human auditory perception.
\end{document}